\documentclass[twocolumn,showpacs,preprintnumbers,amsmath,amssymb,pra]{revtex4}
\usepackage{graphicx}% Include figure files
\usepackage{color}
\begin{document}
%\title{Spectral entanglement and temporal structure of SPDC biphoton states}

\title{Frequency and temporal entanglement of biphoton states in
spontaneous parametric down conversion with a short-pulse pump}

\author{Yu.M. Mikhailova$^{1,\,2}$, P.A. Volkov$^1$,}
\author{M.V. Fedorov$^{1}$}
 \email{fedorov@ran.gpi.ru}
 \affiliation{$^1$A.M.~Prokhorov General Physics Institute,
 Russian Academy of Sciences, Russia\\
 $^2$International Laser Center, Moscow State University, Russia}

\date{\today}

\begin{abstract}
Spectral and temporal coincidence and single-particle photon wave
packets are described and their widths and durations are found.
The degree of entanglement is characterized by the  experimentally
measurable parameter $R$ defined as the ratio of the coincidence
and single-particle spectral widths. In the frequency
representation, this parameter is found as a fnction of the
duration $\tau$ of the pump pulses. The function $R(\tau)$ is
shown to have a minimum and even in the minimum, at rather natural
conditions, the parameter $R$ is found to be very high
$R_{\min}\approx 73\gg 1$. The Schmidt number $K$ is found for
both short and long pump pulses and interpolated for arbitrary
pulse durations. All functional dependences of $R$ and $K$ are
found to be identical and numerical difference between them is
shown to be not exceeding $20\,\%$. Two-time temporal wave
function of a biphoton state is investigated in details, and a
rather significant difference between the cases of short and long
pump pulses is found to occur. In the case of long pulses, the
temporal parameter $R_t$ is defined as the ratio of durations of
the single-particle and coincidence signals, and $R_t$ is shown to
be very close to the Schmidt number $K$.
\end{abstract}


\maketitle

\section{Introduction}
Entanglement is a fundamental feature of multipartite systems
characterizing the degree of correlations between particles.
Entangled states find their wide application in quantum
information, quantum cryptography and related topics. Spontaneous
Parametric Down Conversion (SPDC) is one of the most often used
processes for production of entangled biphoton states. The most
popular and widely discussed type of entanglement of SPDC states
is the entanglement with respect to polarization variables. On the
other hand, in addition to polarization, SPDC photons are
characterized also by other degrees of freedom, such as angles
determining directions of propagation and lengths of wave vectors
(or frequencies) of photons. In contrast to polarization, these
degrees of freedom belong to the class of continuous variables.
Multi- and bipartite states with continuous variables are quite
attractive because they correspond to an infinitely high
dimensionality of the Hilbert space and, potentially, their degree
of entanglement can be very high. But in practice, of course,
there are limitations. Nevertheless, there are conditions when the
degree of entanglement with respect to continuous variables can be
really high, much higher than the maximal achievable degree of
entanglement with respect to discrete variables. For example, in
Refs. \cite{we} SPDC biphoton states were shown to be highly
entangled with respect to angular variables. In this work we
investigate entanglement of SPDC biphotons with respect to their
frequency variables, or spectral entanglement.

We characterize the degree of entanglement by the recently
introduced \cite{2004} parameter $R$ defined as the ratio of
widths of single-particle and coincidence wave packets. We
find $R$ analytically in two asymptotic cases of short and long
pump pulses (with the parameter separating these two regions
explicitly defined) and we show that in these two cases $R$ is
extremely high ($\sim 300$ or more). By using both interpolation
formulas and exact numerical calculations we show that in the case
of the type-I phase matching we consider, $R$ is very high ($\sim
70$) even in the most unfavorable case of intermediately long pump
pulses ($\sim 1\,{\rm ps}$).

In contrast to all other entanglement quantifiers, the parameter
$R$ is rather easy experimentally measurable. For double-Gaussian
bipartite states the parameter $R$ is known \cite{JPB} to coincide
with such a well known entanglement quantifier as the Schmidt
number $K$. However, the wave functions characterizing the
biphoton frequency distributions are rather far from the
double-Gaussian ones. Nevertheless, as we show in this paper, even
for such functions the parameters $R$ and $K$ are very close to
each other. By using specific features of the biphoton wave
functions we calculate in Section ${\bf V}$ the Schmidt number
explicitly in both cases of short and long pump pulses and we find
that in both cases all functional dependences of $R$ and $K$ are
identical, and there is only a small difference between them in
numerical coefficients. This means that the parameter $R$ is,
indeed, an appropriate quantifier of entanglement which can be both
found theoretically and measured experimentally.

In addition to spectral analysis we describe also a temporal
structure of the SPDC biphoton wave packet determined by its
two-time wave function. This structure has many very interesting
features. We derive a series of very simple formulas, which are
used to explain and characterized analytically parameters of the
calculated temporal envelopes of single-particle and coincidence
signals. We find that features of these curves are qualitatively
different in the cases of short and long pump pulses. In the case
of short pump pulses duration of coincidence temporal signals is
shown to be varying in very large limits with varying time of
observation, and this makes the temporal picture with short pump
pulses inappropriate for defining the temporal parameter $R$,
$R_t$. In contrast, in the case of long pump pulses, duration of
coincidence temporal signals is shown to be independent of the
observation time, the parameter $R_t$ is well defined and is shown
to be very close to the Schmidt number $K$.

\section{General formulae}

SPDC is the process in which a classical pump wave propagates
trough a birefringent crystal, where some photons of the pump are
absorbed and give rise to the birth of two photons of smaller
frequencies, signal and idler. If $\bf{k}_1$ and $\bf{k}_2$ are
wave vectors of emitted photons, the QED state arising in the SPDC
process is given by the sum of the vacuum state and superposition
of two-photon states
\begin{equation}
 \label{qed-state}
 |{\rm SPDC}\rangle=|{\rm vac}\rangle
 +\sum_{\bf{k}_1,\,\bf{k}_2}\,\Psi(\bf{k}_1,\,\bf{k}_2)|\bf{k}_1,\,\bf{k}_2\rangle,
\end{equation}
where  $|\bf{k}_1,\,\bf{k}_2\rangle=a_{\bf{k}_1}^\dag
a_{\bf{k}_2}^\dag|{\rm vacuum}\rangle$, $a_{\bf{k}}^\dag$ is the
operator of creation of a photon in a state with the momentum
(wave vector) $\bf{k}$, and $\Psi(\bf{k}_1,\,\bf{k}_2)$ is the
biphoton wave function in the momentum representation. General
expressions for this wave function are well known \cite{Rubin,
Klyshko}, and for pump pulses of a finite duration the biphoton
wave function can be written in the form
\begin{gather}
 \Psi({\bf k}_{1\,\perp},\,{\bf k}_{2\,\perp};
 \,\omega_1,\,\omega_2)=\chi\int d{\bf r}
 \int dt\int d{\bf k}_{p\,\perp}\int d\omega_p\nonumber \\
 \times E_p({\bf k}_{p\,\perp},
 \,\omega_p)\exp[i({\bf k}_{p\,\perp}-{\bf k}_{1\,\perp}-{\bf k}_{2\,\perp})
 \cdot{\bf r}_\perp+i\Delta\cdot z]\nonumber\\
 \times\exp[i(\omega_1+\omega_2-\omega_p)t],\label{Psi-gen}
\end{gather}
where $\chi$ is the crystal susceptibility, indices $p$ and
$\perp$ indicate, correspondingly, the pump and the plane
perpendicular to the laser axis ($z$-axis), $E_p$ is the Fourier
transform of the electric field amplitude of the pump, and
$\Delta$ is the longitudinal mismatch
\begin{equation}
 \label{Delta-gen}
 \Delta=\sqrt{\frac{n_p^2\omega_p^2}{c^2}-{\bf k}_{p\,\perp}^2}
 -\sqrt{\frac{n_1^2\omega_1^2}{c^2}-{\bf k}_{1\,\perp}^2}
 -\sqrt{\frac{n_2^2\omega_2^2}{c^2}-{\bf k}_{2\,\perp}^2},
\end{equation}
 $n_p$, $n_1$, and $n_2$ are the refractive indices of the pump and emitted photons.

Let us make now a series of simplifying assumptions and
approximations. Let us consider the type-I SPDC decay process
$e\rightarrow o+o$, in which both of emitted photons belong to the
ordinary wave and, hence, $n_1=n_2=n_o$. Integration over time $t$
in Eq. (\ref{Psi-gen}) from $-\infty$ to $\infty$ gives
$2\pi\delta(\omega_p-\omega_1-\omega_2)$, with
which the integral over $\omega_p$ can be easily taken. Similarly,
in the wide-crystal approximation the limits of integration over
${\bf r}_\perp$ can be extended to $-\infty$ and $\infty$ in both
transverse directions which gives $(2\pi)^2\delta({\bf
k}_{p\,\perp}-{\bf k}_{1\,\perp}-{\bf k}_{2\,\perp})$, and this
can be used to take the integral over ${\bf k}_{p\,\perp}$. Let us
restrict our consideration in this work only by the case of a
purely collinear propagation of emitted photons, in which ${\bf
k}_{1\,\perp}={\bf k}_{2\,\perp}=0$ and, hence, ${\bf
k}_{p\,\perp}=0$. Besides, let us consider here only the case of
degenerate central frequencies $\omega_1^{(0)}$ and
$\omega_2^{(0)}$, $\omega_1^{(0)}=\omega_2^{(0)} =\omega_0/2$,
where $\omega_0$ is the central frequency of the pump. We will
assume also that deviations from these central frequencies,
$\nu_{1,\,2}\equiv\omega_{1,\,2} -\omega_{1,\,2}^{(0)}$, are
small: $|\nu_{1,\,2}|\ll\omega_0$. Finally, we will take the
spectral amplitude of the pump in a Gaussian form
\begin{gather}
 E_p({\bf k}_{p\,\perp}=0,\,\omega_p)\equiv E_p(\omega_p)=
 E_0\exp\left[-\frac{(\omega_p-\omega_0)^2\tau^2}{8\ln 2}\right]\nonumber\\
 = E_0\exp\left[-\frac{(\omega_1+\omega_2-\omega_0)^2\tau^2}{8\ln 2}\right]\label{Gauss},
\end{gather}
where $\tau$ is the duration of pump pulses and
$\omega_p=\omega_1+\omega_2=\omega_0+\nu_1+\nu_2$.

Under the formulated assumptions and approximations  Eq.
(\ref{Psi-gen}) for the biphoton wave function is reduced to a
much simpler form
\begin{equation}
 \label{Freq-wf-int-dz}
 \Psi(\omega_1,\,\omega_2)\propto E_p(\omega_1+\omega_2)\,
 \int_0^L dz e^{i\,\Delta (\omega_1,\,\omega_2) z}
\end{equation}
or, after integration over $z$,
\begin{equation}
 \label{Freq-wave-function}
 \Psi(\omega_1,\,\omega_2)\propto E_p(\omega_1+\omega_2)\,
 {\rm sinc} \left[\frac{L}{2}\Delta(\omega_1,\,\omega_2)\right],
\end{equation}
where ${\rm sinc}(u)=\sin u/u$, $L$ is the length of the crystal
(along the laser axis), and the phase mismatch
$\Delta(\omega_1,\,\omega_2)$ is determined by the equation
following from Eq. (\ref{Delta-gen})
\begin{gather}
 \Delta=k_p-k_1-k_2\nonumber\\
 =\frac{(\omega_1+\omega_2)
 n_p(\omega_1+\omega_2)-\omega_1 n_0(\omega_1)-\omega_2 n_0(\omega_2)}{c}.\label{mismatch}
\end{gather}
Note, that as we will be interested mainly in shapes  and widths
of photon spectral distributions, in Eqs. (\ref{Freq-wf-int-dz})
and (\ref{Freq-wave-function}) and henceforth we drop all the
constant coefficients on the right-hand sides of such equations by
using the proportionality symbol. If not specifies differently, all
curves describing spectral or temporal distributions will be normalized
by the condition of being equal to unity at maxima.

Owing to the assumption about small deviations of frequencies
$\omega_{1,\,2}$ from $\omega_{1,\,2}^{(0)}$, the expression on
the right-hand side of Eq. (\ref{mismatch}) can be expanded in
powers of $\nu_1$ and $\nu_2$ with only two lowest orders to be
taken into account
\begin{equation}
 \label{mismatch-approx}
 \Delta\approx \frac{1}{c}\left[A(\nu_1+\nu_2)-2B\frac{\nu_1^2+\nu_2^2}{\omega_0}\right],
\end{equation}
where $A$ is the dimensionless constant characterizing the
temporal walk-off
\begin{gather}
 A=c\left(\left.k_p^\prime(\omega)\right|_{\omega=\omega_0}-
 \left.k_1^\prime(\omega)\right|_{\omega=\omega_0/2}\right)\nonumber\\
 =c\left(\frac{1}{v_g^{(p)}}-\frac{1}{v_g^{(o)}}\right)\label{A};
\end{gather}
$v_g^{(p)}$ and $v_g^{(o)}$ are the group velocities of the pump
and ordinary waves ($v_g^{(o)}>v_g^{(p)}$); $B$ in Eq.
(\ref{mismatch-approx}) is the dispersion constant, also
dimensionless
\begin{equation}
 \label{B}
 B=\frac{c}{4}\,\omega_0\left.k_1^{\prime\prime}(\omega)\right|_{\omega=\omega_0/2}.
\end{equation}
For ${\rm LiIO_3}$ crystal and $\lambda_0=400\,{\rm nm}$ numerical
values of the constants $A$ and $B$ are: $A=0.17$ and $B=0.069$.
Though small, nevertheless, the quadratic term in Eq.
(\ref{mismatch-approx}) is crucially important for determining
finite width of the photon single-particle distribution (see the
next section, part ${\rm \bf{B}}$). Note also that in Eq.
(\ref{mismatch-approx}) we have dropped the dispersion arising
from the wave vector of the pump and proportional to
$k_1^{\prime\prime}(\nu_1+\nu_2)^2$. This term is dropped because
it depends on the same combination of frequencies as the linear
term $\nu_1+\nu_2$ and determines only a small correction to the
linear term not changing qualitatively its functional dependence
on $\nu_1$ and $\nu_2$. In the same approximation we can further
simplify the dispersion  term taken into account in Eq.
(\ref{mismatch-approx}):
$\nu_1^2+\nu_2^2\equiv\frac{1}{2}[(\nu_1+\nu_2)^2]+(\nu_1-\nu_2)^2]\rightarrow
\frac{1}{2}(\nu_1-\nu_2)^2$. With this substitution the final
expression for the biphoton wave function takes the form
\begin{gather}
 \notag
 \Psi(\nu_1,\,\nu_2)\propto\exp\left(-\frac{(\nu_1+\nu_2)^2\tau^2}{8\ln
2}\right)\\
\times{\rm
 sinc}\left\{\frac{L}{2c}\left[A(\nu_1+\nu_2)-B\frac{(\nu_1-\nu_2)^2}{\omega_0}\right]
 \right\}.\label{Freq-wave-function-2}
\end{gather}

\section{Coincidence and single-particle distributions (short pulses)}

\subsection{Coincidence spectrum}
\begin{figure}[ht]
\centering\includegraphics[width=8.5cm]{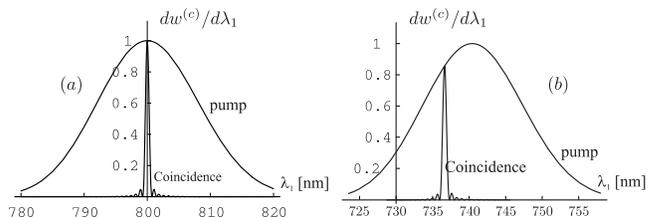}
\caption{{\protect\footnotesize {Coincidence and pump spectra
(narrow and wide curves) as functions of $\lambda_1$ at
$\lambda_2=\,(a)\, 800\,{\rm nm}$ and $(b)\,870\,{\rm nm}$.
}}}\label{Fig1}
\end{figure}

The coincidence spectrum is determined by the squared wave
function $|\Psi(\nu_1,\,\nu_2)|^2$ of Eq.
(\ref{Freq-wave-function-2}) at a given value of $\nu_2$. For
practical purposes, it is more convenient to consider the
coincidence and single-particle probability densities as functions
of wavelengths $\lambda_{1,\,2}$ rather than frequencies
$\nu_{1,\,2} =2\pi c/\lambda_{1,\,2}-\pi c/\lambda_0$. For the
crystal length and pump-pulse duration equal to $L=0.5\,{\rm cm}$
and $\tau=50\,{\rm fs}$, the coincidence spectral distribution
$dw^{(c)}/d\lambda_1$ is characterized by narrow curves of Fig.
\ref{Fig1}. The wide curves in the pictures of Fig. \ref{Fig1}
describe the pump spectrum. The curves of the pictures $(a)$ and
$(b)$ are calculated at  the second-photon wavelength $\lambda_2$
taken equal to  $800\,{\rm nm}$ and $870\,{\rm nm}$.

Fig. \ref{Fig1} clearly shows  that ($i$) the coincidence spectrum
is much narrower than that of the pump and ($ii$) the coincidence
spectral width $\Delta\lambda_1^{(c)}$ does not depend on
$\lambda_2$. If we take the wavelength of the second photon
exactly equal to the double pump wavelength,
$\lambda_2=2\lambda_0=800\,{\rm nm}$, the peaks of the pump and
coincidence spectra coincide and are located at
$\lambda_1=800\,{\rm nm}$ (Fig. \ref{Fig1}$(a)$)). If we shift
$\lambda_2$ to the right, both pump and coincidence spectra move
to the left from the point $\lambda_2=800\,{\rm nm}$. However, for
the coincidence spectral distribution this shift is slightly
larger than for the pump. For this reason the amplitude of the
coincidence signal appears to be somewhat smaller than in the case
of location at $800\,{\rm nm}$. However, the width of the
coincidence spectral distribution remains practically the same as
in the case $\lambda_2=800\,{\rm nm}$. Hence, the coincidence
spectral width $\Delta\lambda_1^{(c)}$ is a constant parameter,
independent of tuning of the second-photon detector.

Analytically the coincidence spectral width can be easily found
directly from Eq. (\ref{Freq-wave-function-2}). In the case of
short pulses the Gaussian function in this equation is much
narrower than the sinc-function. For bipartite wave functions
having the form of a product of two functions, one of which is
much narrower than another, the coincidence distribution is
determined by the narrower factor, i.e. in our case by the
sinc-function. Evidently, it's width with respect to varying
$\nu_1$ at a given value of $\nu_2$ is given by
\begin{equation}
 \label{coi-fwhm}
 \Delta\nu_1^{(c)}=\Delta\nu_{1\,{\rm sinc}}=\frac{2.78\times\,2c}{AL},
\end{equation}
where the factor 2.78 is the full width at half-maximum (FWHM) of
the function ${\rm sinc}^2(u)$. Numerically, at $L=0.5\,{\rm cm}$,
in terms of wavelengths, both analytical expression and
calculations shown in Fig. \ref{Fig1} correspond to
$\Delta\lambda_1^{(c)}= 0.658\,{\rm nm}$. For comparison, FWHM of
the pump spectral function $\exp[-(\nu_1+\nu_2)^2\tau^2/4\ln 2]$
at a given $\nu_2$ equals to
\begin{equation}
 \label{fwhm-exp}
 \Delta\nu_{1\,{\rm pump}}=\frac{4\ln 2}{\tau},
\end{equation}
which gives (in terms of wavelengths, at $\tau=50\,{\rm fs}$ and
$\lambda_0=400\,{\rm nm}$)
$\Delta\lambda_1^{(p)}=2\lambda_0^2\Delta\nu_{1\,{\rm pump}}/\pi
c=18.8\,{\rm nm}$. In other words, the pump spectrum considered as
a function of $\lambda_1$,
\begin{equation}
 \notag
 E_p^2\propto\exp\left\{-\frac{\pi^2 c^2\tau^2}{\ln 2}
 \left(\frac{1}{\lambda_1}+\frac{1}{\lambda_2}-\frac{1}{\lambda_0}\right)^2\right\},
\end{equation}
is 28.57 times wider than the coincidence spectrum.

As shown below, the ratio of the ``sinc" and ``pump" widths
[(\ref{coi-fwhm}) and (\ref{fwhm-exp})] determines the parameter
separating the regions of short and long pulses
\begin{equation}
 \label{eta}
 \eta=\frac{\Delta\nu_{1\,{\rm sinc}}}{\Delta\nu_{1\,{\rm pump}}}
 \approx 2\frac{c\,\tau}{AL}=\frac{2\tau}{L/v_g^{(p)}-L/v_g^{(o)}}.
\end{equation}
The last expression on the right-hand side of this equation shows
that $\eta$ equals the ratio of the double pump-pulse duration to
the difference of times during which a crystal is crossed by the
pump and idler/signal photons. Also $\eta$ can be considered as
the dimensionless pump-pulse duration. In SPDC the pump pulses are
short if $\eta\ll 1$ and long if $\eta\gg 1$. The boundary pulse
duration occurs at $\eta\sim 1$. At $L=0.5\,{\rm cm}$ the equality
$\eta=1$ yields $\tau=1.43\,{\rm ps}$. As shown below, $\eta$ is
also the key control parameter, determining a value of the
entanglement parameter $R$.

\begin{figure}[ht]
\centering\includegraphics[width=6.5cm]{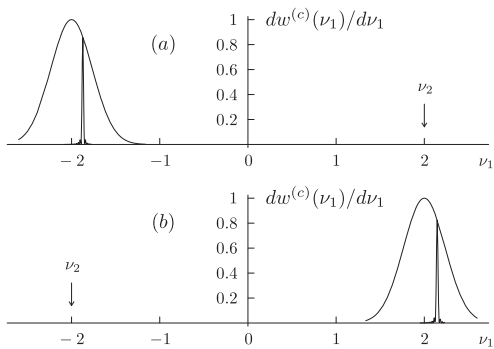}
\caption{{\protect\footnotesize {Coincidence and pump spectra
$dw^{(c)}(\nu_1)/d\nu_1$ at $(a)\,\nu_2=2$ and $(b)\,\nu_2=-2$,
$\nu_1$ and $\nu_2$ are in units of $10\,{\rm fs}$.
}}}\label{Fig2}
\end{figure}

Non-coincidence of the peaks of the pump and coincidence spectra
seen in Fig. \ref{Fig1}$\,(b)$ deserves a special comment. For
clarification of its origin, it's worth to return to the frequency
pictures. In Fig. \ref{Fig2} the coincidence and pump spectra are
shown as functions of $\nu_1$ under the conditions when the
detector registering the second photon of the pair (``2") is tuned
to frequencies shifted as compared to $\omega_0$ (or $\nu_2$
shifted from zero) $(a)$ to the blue  and $(b)$ to the red sides.
Both the pump and coincidence spectra also appear to be shifted
from $\omega_0$ ($\nu_1=0$),  but in the directions opposite to
the shift of $\omega_2$. As is seen, the peaks of the pump and
coincidence spectra do not coincide at sufficiently large values
of these shifts. Interesting enough, in both cases the coincidence
peak appears to be located at the right wing of the pump, which
corresponds to the left wing in the wave-length picture of Fig.
\ref{Fig1}. This lack of symmetry between the pictures $(a)$ and
$(b)$ of Fig. \ref{Fig2} is related to the dispersion and, as we
will see in the following section,  it is related to asymmetry of
the single-particle spectrum. Mathematically the results discussed
here follow directly from the definition of the phase-mismatch
$\Delta$, e.g., taken in the form (\ref{mismatch-approx}).
Location of the coincidence spectral peaks is determined by the
condition $\Delta=0$. This gives a quadratic equation of $\nu_1$,
solution of which is given by
\begin{gather}
 \notag
 \nu_1(\nu_2)=\frac{A\omega_0}{4B}-\sqrt{\left(\frac{A\omega_0}{4B}\right)^2
 +\frac{A\omega_0}{2B}\nu_2-\nu_2^2}\\
 \label{nu-1(nu-2)}
 \approx -\nu_2+\frac{4B\,\nu_2^2}{A\,\omega_0}.
\end{gather}
The first and second terms on the right-hand side of this equation
have different parity. They are, correspondingly, odd and even
functions of $\nu_2$, and this explains the lack symmetry between
the pictures $(a)$ and $(b)$ of Fig. \ref{Fig2}.

The approximation used in the expansion of the square root in Eq.
(\ref{nu-1(nu-2)}) consists in the assumption $|\nu_2|\ll
A\omega_0/8B$, which is satisfied for parameters we use in this
paper. Actually, if here the approximate expression for
$\nu_1(\nu_2)$ was needed only for getting a simple analytical
formula explaining the exact calculations of the spectra in Figs.
\ref{Fig1} and \ref{Fig2}, later this will be shown to be
crucially important, in particular, for calculations of the
Schmidt number (Section ${\bf V})$.

\subsection{Single-particle spectrum and the entanglement parameter for short pump pulses}

\begin{figure}[h]
\centering\includegraphics[width=6.5cm]{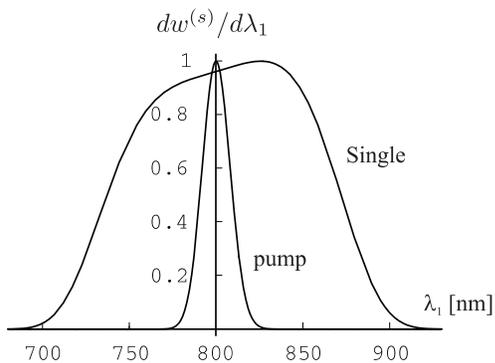}
\caption{{\protect\footnotesize {Single-particle (wide) and pump
(narrow) spectra (densities of probability of the corresponding
photon distributions). }}}\label{Fig3}
\end{figure}

The single-particle photon spectrum is determined by the squared
wave function of Eq. (\ref{Freq-wave-function-2})
integrated over $\nu_2$
\begin{equation}
 \label{single}
 \frac{dw^{(s)}}{d\nu_1}=\int d\nu_2\,|\Psi(\nu_1,\,\nu_2)|^2.
\end{equation}
As the exact analytical calculation of the integral over $\nu_2$
is hardly possible, let us perform this integration approximately
with the narrow sinc$^2$-function substituted by the
$\delta$-function $\delta[\nu-\nu(\nu_1)]$, where
$\nu=\nu_1+\nu_2$ and $\nu(\nu_1)$ is given by equation similar to Eq. (\ref{nu-1(nu-2)})
\begin{gather}
 \notag
 \nu(\nu_1)=\frac{A\omega_0+4B\nu_1-\sqrt{A^2\omega_0^2+8AB\omega_0\nu_1}}{2B}\\
 \label{nu}
 \approx\frac{4B\nu_1^2}{A\omega_0}\,.
\end{gather}
By using now $\nu$ as the integration variable in Eq.
(\ref{single})  (instead of $\nu_2$), we find immediately the
following expression for the single-particle spectrum of emitted
photons
\begin{equation}
 \label{Freq-Single}
 \frac{dw^{(s)}(\nu_1)}{d\nu_1}\propto \frac{\exp\left\{\displaystyle
 -\frac{[\nu(\nu_1)]^2\tau^2}{4\ln 2}\right\}}
 {\sqrt{A^2\omega_0^2+8AB\omega_0\nu_1}}.
\end{equation}
In Fig. \ref{Fig3} the spectral distribution determined by  Eq.
(\ref{Freq-Single}) is shown together with the pump spectrum
(correspondingly, the wide and narrow curves).
The FWHM of the wide curve in Fig. \ref{Fig3} equals to
\begin{equation}
 \label{width-single}
 \Delta\nu_1^{(s)}=\Delta\omega_1^{(s)}=\sqrt{\frac{2A\ln (2)\,\omega_0}{B\tau}}.
\end{equation}
Numerically, at $\tau=50\,{\rm fs}$, Eq. (\ref{width-single})
yields $\Delta\lambda_1^{(s)}=195\,{\rm nm}$, which is 10 time
larger than the pump spectral width.

The parameter $R$ defined as the ratio of the single-particle to
coincidence  spectral widths is easily found from Eqs.
(\ref{coi-fwhm}) and (\ref{width-single}) to be given by
\begin{gather}
 \notag
 R_{\rm short}= \frac{\Delta\omega_1^{(s)}}{\Delta\omega_1^{(c)}}
 =\frac{A^{3/2}}{2.78}\sqrt{\frac{\pi\ln{2}}{B}}\frac{L}{\sqrt{\lambda_0\, c\tau}}=\\
\frac{\sqrt{2\pi\ln
 2}}{2.78}\frac{A}{\sqrt{B\,\eta}}\sqrt{\frac{L}{\lambda_0}}
 =0.7507\frac{A}{\sqrt{B\,\eta}}\sqrt{\frac{L}{\lambda_0}}
 =\frac{55.06}{\sqrt{\eta}},\label{ratio}
\end{gather}
where the subscript $``{\rm short}"$ emphasizes that this result
is valid only for short pump pulses, $\eta\ll 1$, and  the control
parameter $\eta$ is defined in Eq. (\ref{eta}). The numerical
coefficient 55.06 in the last expression of Eq. (\ref{ratio}) was
found at the same values of all parameters which were used
earlier, except the pump-pulse duration $\tau$ which was not fixed
yet. At $\tau=50\,{\rm fs}$ ($\eta=0.0348$) Eq. (\ref{ratio})
yields
\begin{equation}
 \label{R-short}
 R_{\rm short}\approx 295.
\end{equation}
This degree of entanglement is really high. As it's seen from the
given derivation, such an extremely high degree of entanglement is
related to two accompanying effects: narrowing of the coincidence
and broadening of the single-particle spectra of SPRC photons
(compared to the pump spectrum). The second of these two effects
is a specific feature of the type-I SPDC process. Indeed, in the
case of the type-II pase matching ($o\rightarrow e+o$), in the
linear approximation, the expansion of the phase mismatch in
powers of $\nu_1$, $\nu_2$ takes the form
\begin{equation}
 \label{type-II}
 \Delta=\left(k_p^\prime-\frac{k_1^\prime+k_2^\prime}{2}\right)(\nu_1+\nu_2)
 -\frac{k_1^\prime-k_2^\prime}{2}(\nu_1-\nu_2).
\end{equation}
As in the case of the type-II phase matching the emitted photons
``1" and ``2" are  different ($o$- and $e$-waves), derivatives of
their wave vectors over frequencies are different too,
$k_1^\prime\neq k_2^\prime$. For this reason, the phase mismatch
of Eq. (\ref{type-II}) already in the linear approximation appears
to be depending on both the sum and difference of frequencies,
$\nu_1+\nu_2$ and $\nu_1-\nu_2$. In this case the linear
approximation in $\nu_1$, $\nu_2$ is sufficient for determining
finite widths of both coincidence and single-particle
distributions. If $k_1^\prime-k_2^\prime$ is of the same order as
$k_1^\prime+k_2^\prime$, the difference between the
single-particle widths and that of the pump spectrum is not too
large. I.e., the effect of broadening of the single-particle
distribution is missing or is only weakly pronounced. In contrast
to this, in the case of the type-I phase matching both emitted
photons belong to the same ordinary type of a wave and, hence,
$k_1^\prime=k_2^\prime$. In the linear approximation the
dependence of the phase mismatch $\Delta$ (\ref{type-II}) on the
difference of frequencies $\nu_1-\nu_2$ disappears, and the linear
approximation appears to be insufficient for determining the finite
width of the single-particle distribution. To define this width
one has to take into account the second order (dispersion) term in
the expansion of $\Delta$ in powers of $\nu_1$, $\nu_2$. As this
term is rather small, the width of the single-particle
distribution appears to be rather large. I.e., equality of
$k_1^\prime$ and $k_2^\prime$ gives rise to the above-described
broadening of the single-particle spectrum compared to that of the
pump. This effect provides at least the 10-fold increase of the
parameter $R$ compared to what can be achieved in the type-II SPDC
process.

Note that the above-described difference between the type-I and
type-II SPDC processes was understood rather long ago
\cite{Keller-Rubin}). New elements of our consideration consist in
demonstration of importance of this effect for producing byphoton
states with extremely high degree of entanglement and in explicit
evaluation of the achievable degree of entanglement. As far as we
know, there was only one experiment where both coincidence and
single-particle SPDC spectra were measured simultaneously
\cite{Kurts}. However, the difference between the coincidence and
single-particle distributions observed in this work was not too
high and corresponded to the ratio of widths on the order of 2. We
assume that one of the reasons why this ratio was so small is in
the type of phase-matching (type-II).

In Ref.\cite{Law-Eberly} the degree of spectral entanglement for
the type-II SPDC process was evaluated with the help of
decomposition of the wave function in a series of products of
Schmidt modes, which is different from our approach.

In some works (see, e.g., Ref. \cite{Christine}) the 2D photon
distribution was measured and plotted in the $(\omega_1,\,
\omega_2)$ plane. However, as far as we know, such results were
never used to evaluate quantitatively the degree of entanglement.

Finally, in this work we discuss only possibilities of direct
measurements of the coincidence and single-particle spectral and
temporal distributions of photons. We do not discuss here
connections between results obtained in such a way with
interference measurements, e.g., in Refs. \cite{Walmsley,
Sergienko,Grice,Chekhova,Kim}. We hope to return to this problem
elsewhere.

\section{Long pump pulses and interpolation for arbitrary pulse durations}
Until now we have assumed that the sinc-function in Eq.
(\ref{Freq-wave-function-2}) is much narrower than the Gaussian
function. This is absolutely true, e.g., for the pulse duration
$\tau=50\,{\rm fs}$, which was chosen for estimates and
illustrations. However, if the pulse duration is significantly
longer, the relation between the two functions on the right-hand
side of Eq. (\ref{Freq-wave-function-2}) may be reversed: the
Gaussian function can become much narrower than the sinc-function.
In this case the coincidence distribution is determined by the
pump spectrum. E.g., for $\nu_2=0$ it has the form
\begin{equation}
 \label{coi-distr-long}
 \frac{dw^{(c)}}{d\nu_1}\propto \exp\left[-\frac{\nu_1^2\tau^2}
 {4\ln 2}\right].
\end{equation}
In the integral (\ref{single}) determining the single-particle
probability density the narrow pump spectrum can be approximated by
$\delta(\nu_2+\nu_2)$ to give
\begin{equation}
 \label{single-distr-long}
 \frac{dw^{(s)}}{d\nu_1}\propto \left\{ {\rm sinc}\left[\frac{2LB\nu_1^2}{c\,\omega_0}\right]\right\}^2.
\end{equation}
The coincidence and single-particle width found from Eqs.
(\ref{coi-distr-long}) and (\ref{single-distr-long}) are equal to
\begin{equation}
 \label{width-long-tau}
 \Delta\nu_{1\,{\rm long}}^{(c)}=\frac{4\ln 2}{\tau},\quad
 \Delta\nu_{1\,{\rm long}}^{(s)}=\sqrt{\frac{2.78\,c\,\omega_0}{L\,B}},
\end{equation}
and the corresponding entanglement parameter $R$ is given by
\begin{gather}
 \notag
 R_{\rm long}=\frac{\Delta\nu_{1\,{\rm long}}^{(s)}}
 {\Delta\nu_{1\,{\rm long}}^{(c)}}=\frac{\sqrt{2.78\,\pi}}{2^{3/2}\ln
 2}\,\frac{c\tau}{\sqrt{BL\lambda_0}}=\\
 \label{R-long}
 =\frac{\sqrt{2.78\,\pi}}{2^{5/2}\ln
 2}\frac{A\eta}{\sqrt{B}}\sqrt{\frac{L}{\lambda_0}}
 =0.7537\frac{A\eta}{\sqrt{B}}\sqrt{\frac{L}{\lambda_0}}
 =55.28\,\eta.
\end{gather}
The entanglement parameter $R_{\rm long}$ (\ref{R-long}) is a
linearly growing function of the pulse duration $\tau$. E.g., at
$\tau = 7 {\rm ps}$ Eq. (\ref{R-long}) yields $\eta\approx 4.87$
and $R_{\rm long}\approx 269\gg 1$.

\begin{figure}[h]
\centering\includegraphics[width=6cm]{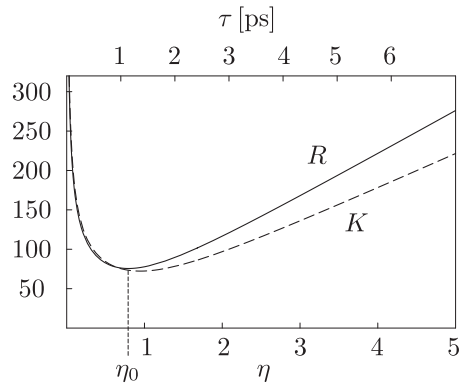}
\caption{{\protect\footnotesize {The parameter $R$ and $K$ vs. the
control parameter $\eta$ (\ref{eta}). }}}\label{Fig4}
\end{figure}

For evaluating the entanglement parameter in the whole range of variation of the pump-pulse duration $\tau$ (from
short- to long-pulse regions) we use the simplest interpolation
formula
\begin{equation}
 \label{interpolation}
 R(\tau)= \sqrt{R_{\rm short}^2(\tau)+R_{\rm long}^2(\tau)}.
\end{equation}
Note that there is an amusing, though approximate, coincidence
relation between numerical coefficients in Eqs. (\ref{ratio}) and
(\ref{R-long}): $\sqrt{2\pi\ln 2}/2.78\approx
1.004\,\sqrt{2.78\pi}/(2^{5/2}\ln 2).$ Hence, with a very good
accuracy we can take coefficients in Eqs. (\ref{ratio}) and
(\ref{R-long}) coinciding with each other to get
\begin{equation}
 \label{overall}
 R(\eta)=0.75\frac{A}{\sqrt{B}}\sqrt{\frac{L}{\lambda_0}}
 \sqrt{\eta^2+\frac{1}{\eta}}=55\sqrt{\eta^2+\frac{1}{\eta}}\,.
\end{equation}
The function $R(\eta)$ is plotted in Fig. \ref{Fig4}.
The curve has a minimum at $\eta_0=2^{-1/3}=0.79$, which
corresponds to $\tau_0=1.14\,{\rm ps}$, and in the minimum
\begin{equation}
 \label{R-min}
 R_{\min}=R(\tau_0)\approx\frac{A}{\sqrt{B}}\sqrt{\frac{L}{\lambda_0}}\approx 73\gg 1.
\end{equation}

So, the parameter $R(\tau)$ is very large and the degree of
entanglement is very high in both earlier considered cases of
short and long pump pulses, $\tau\ll\tau_0$ and $\tau\gg\tau_0$.
But $R(\tau)$ is large and entanglement is high also in the most
unfavorable case of intermediate pump-pulse durations around
$\tau_0$. Actually, the value of $R(\tau)$ around its minimum is
determined by the well defined parameter equal to the square root
of the ratio "length of the crystal divided by the pump
wavelength". This result shows that  in the type-I degenerate
collinear SPDC process, practically, {\it spectral entanglement is
never small}.

\section{The Schmidt number}

As mentioned above (in Introduction), in the case of
double-Gaussian wave functions, the parameter $R$ coincides
identically with the Schmidt number $K$ \cite{JPB}, which is
widely recognized to be a good quantifier of the degree of
entanglement in pure bipartite states. However, the wave function
of the form (\ref{Freq-wave-function-2}) is not double-Gaussian.
Actually, if the pump spectral amplitude can be taken Gaussian,
the sinc-function with the argument containing both linear and
quadratic terms in variables $\nu_{1,\,2}$ cannot be even modeled
by any Gaussian form. For this reason, the question that often
arises is whether the parameter $R$ can be used as the
entanglement quantifier for states characterized by such
non-double-Gaussian wave functions, while relation between $R$ and
$K$ remains unknown. The question is reasonable and, to be
answered, it requires the Schmidt number to be explicitly
calculated. Solution of this problem is given below.

Let us use the well known general integral definition of the
Schmidt number for bipartite systems with continuous variables
[Eq. (10) of Ref. \cite{JPB}]. In the case of spectral
entanglement this definition takes the form
\begin{gather}
 \notag
 K=N^2\Bigg[\int d\nu_1 d\nu_2 d\nu_1^\prime d\nu_2^\prime\Psi(\nu_1,\,\nu_2)
 \Psi(\nu_1,\,\nu_2^\prime)\\
 \label{K-general}
 \times\Psi(\nu_1^\prime,\,\nu_2)\Psi(\nu_1^\prime,\,\nu_2^\prime)\Bigg]^{-1},
\end{gather}
where $N$ is the norm of the wave function $\Psi$
\begin{equation}
 \label{norm}
 N=\int d\nu_1 d\nu_2\,\left|\Psi(\nu_1,\,\nu_2)\right|^2
\end{equation}
and $\Psi(\nu_1,\,\nu_2)$ is given by Eq.
(\ref{Freq-wave-function-2}).

In a general case, the integrals in Eqs. (\ref{K-general}),
(\ref{norm}) cannot be calculated analytically, and even their
numerical calculation is a rather difficult problem. To perform
integrations we will use approximations following from the main
features of the wave function (\ref{Freq-wave-function-2})
described above. Let us consider separately the cases of short and
long pulses, $\eta\ll 1$ and $\eta\gg 1$.

1) {\bf Short pump pulses}.

The mismatch $\Delta(\nu_1,\,\nu_2)$ (\ref{mismatch-approx})
entering the argument of the sinc-function in the definition of
$\Psi(\nu_1,\,\nu_2)$ is quadratic in $\nu_1$. Hence, the equation
$\Delta=0$ has two solutions, $\nu_1^+(\nu_2)$ and
$\nu_1^-(\nu_2)$. Only one of these solutions is compatible with
the condition $|\nu_{1,\,2}|\ll\omega_0$, and this solution is
given by equation similar to Eq. (\ref{nu-1(nu-2)})
\begin{equation}
 \notag
 \nu_1^-(\nu_2)\equiv f(\nu_2)=-\nu_2+\frac{4B}{A\omega_0}\,\nu_2^2.
\end{equation}
In the vicinity of $\nu_1^-$ the mismatch $\Delta(\nu_1,\,\nu_2)$
can be approximated by the linear function of $\nu_1$
\begin{equation}
 \label{Delta-linear}
 \Delta\equiv\Delta (\nu_1,\,\nu_2)\approx\frac{A}{c}\,\big[\nu_1-f(\nu_2)\big].
\end{equation}
The same approximation can be applied to all other functions
$\Psi$ in the definitions of the Schmidt number (\ref{K-general})
and norm of the wave function (\ref{norm}), which take then the
form
\begin{gather}
 \notag
 K^{-1}N^2=\int d\nu_1 d\nu_2 d\nu_1^\prime d\nu_2^\prime\\
 \notag
 \times\exp\left[-\frac{(\nu_1+\nu_2)^2\tau^2}{8\ln 2}\right]
 {\rm sinc}\left[\frac{LA}{2c}\,\Big(\nu_1-f(\nu_2)\Big)\right]\\
 \notag
 \times\exp\left[-\frac{(\nu_1+\nu_2^\prime)^2\tau^2}{8\ln 2}\right]
 {\rm sinc}\left[\frac{LA}{2c}\,\Big(\nu_1-f(\nu_2^\prime)\Big)\right]\\
 \notag
 \times\exp\left[-\frac{(\nu_1^\prime+\nu_2)^2\tau^2}{8\ln 2}\right]
 {\rm sinc}\left[\frac{LA}{2c}\,\Big(\nu_1^\prime-f(\nu_2)\Big)\right]\\
  \label{K-linear}
 \times\exp\left[-\frac{(\nu_1^\prime+\nu_2^\prime)^2\tau^2}{8\ln 2}\right]
 {\rm sinc}\left[\frac{LA}{2c}\,\Big(\nu_1^\prime-f(\nu_2^\prime)\Big)\right]
\end{gather}
and
\begin{gather}
 \notag
 N=\int d\nu_1 d\nu_2\,\exp\left[-\frac{(\nu_1+\nu_2)^2\tau^2}{4\ln
 2}\right]\\
 \label{norm-linear}
 \times\left\{{\rm  sinc}
 \left[\frac{LA}{2c}\,\Big(\nu_1-f(\nu_2)\Big)\right]\right\}^2.
\end{gather}
Still, the integrals (\ref{K-linear}) and (\ref{norm-linear}) are
too complicated for analytical calculation. But we can remember
now that in the case of short pulses ($c\tau\ll AL/2$) the
exponential functions in all integrals are much wider than the
${\rm sinc}$-functions. Hence, with a good accuracy, we can take
off the exponential functions from the integrals at values of
their arguments determined by the condition that the corresponding
phase mismatch turns zero. With such a procedure applied to Eq.
(\ref{norm-linear}) we get
\begin{gather}
 \notag
 N=\frac{2\pi c}{LA}\int d\nu_2\,\exp\left[-\frac{4B^2\nu_2^4\tau^2}{A^2\omega_0^2\ln
 2}\right]\\
 =\frac{2^{3/2}\pi c}{L}(\ln 2)^{1/4}\Gamma\left(\frac{5}{4}\right)\sqrt{\frac{\omega_0}{AB\tau}}
 \label{noorm-final},
\end{gather}
where we took into account that $\int_{-\infty}^\infty {\rm
sinc}^2(x)dx=\pi$ and  $\int_{-\infty}^\infty
e^{-x^4}dx=2\Gamma(5/4)$, $\Gamma$ denotes the gamma-function, and
$\Gamma(5/4)=0.9064$.

For calculation of integrals in Eq. (\ref{K-linear}) we use the
following exact equation for the integrated product of two ${\rm
sinc}$-functions
\begin{equation}
 \label{sinc-sinc}
 \int_{-\infty}^\infty dx\,{\rm sinc}(x)\,{\rm sinc}(x+y)=\pi\,{\rm
 sinc}(y).
\end{equation}
Known or not, this equality can be easily proved, e.g., by means
of analytical continuation to the complex plane $x$ and
integration by the residue method. With the help of the rule
(\ref{sinc-sinc}), we perform integration in Eq. (\ref{K-linear})
separately in the first and second pairs of lines under the symbol
of integrals (correspondingly, over $\nu_1$ and $\nu_1^\prime$),
with the exponential functions taken out of the integrals, as
explained above. As a result we get
\begin{gather}
 \notag
 K^{-1}N^2=\left(\frac{2\pi c}{LA}\right)^2\int d\nu_2 d\nu_2^\prime
 \exp\left[-\frac{4B^2\tau^2}{A^2\omega_0^2\ln
 2}(\nu_2^4+\nu_2^{\prime\,4})\right]\\
 \label{K-intermediate}
 \times\left\{{\rm sinc}\left[\frac{LA}{2c}(\nu_2-\nu_2^\prime)\right]\right\}^2.
\end{gather}
Note that in the last expression the difference
$\nu_1(\nu_2)-\nu_1(\nu_2^\prime)$ in the argument of the ${\rm
sinc}$-function is approximated by $\nu_2^\prime-\nu_2$, i.e.,
here the quadratic term of the expression in Eq,
(\ref{nu-1(nu-2)}) is dropped both in $f(\nu_2)$ and
$f(\nu_2^\prime)$. This is possible because in Eq.
(\ref{K-intermediate}) the argument of the exponential function
differs from that of the ${\rm sinc}$-function, and hence, small
quadratic terms in the ${\rm sinc}$-function argument are not
needed for making integrals converging, as this was, e.g., in the
case of integrations over $\nu_1$ and $\nu_1^\prime$. The
integrals in Eq. (\ref{K-intermediate}) are easily calculated with
the help of the same procedure as used above, with the slow
exponential function taken off from the integral over
$\nu_2^\prime$ at $\nu_2^\prime=\nu_2$. The results are given by
\begin{gather}
 \notag
 K^{-1}N^2=\left(\frac{2\pi c}{LA}\right)^3\int d\nu_2 \exp\left[-\frac{8B^2\tau^2}{A^2\omega_0^2\ln
 2}\nu_2^4\right]\\
 \label{K-semifinal}
 =\left(\frac{2\pi c}{LA}\right)^3
 \Gamma\left(\frac{5}{4}\right)\left(2\ln 2\right)^{1/4}\sqrt{\frac{A\omega_0}{B\tau}}
\end{gather}
and
\begin{gather}
 \notag
 K_{\rm short}=\frac{N^2}{K^{-1}N^2}=\frac{(2\ln 2)^{1/4}}{\sqrt{\pi}}\Gamma\left(\frac{5}{4}\right)
 \frac{A^{3/2}}{\sqrt{B}}\frac{L}{\sqrt{c\tau\,\lambda_0}}\\
 \label{K-final}
 =0.785\,\frac{A}{\sqrt{B\,\eta}}\sqrt{\frac{L}{\lambda_0}}
 =\frac{57.5}{\sqrt{\eta}}.
\end{gather}
Comparison with Eq. (\ref{ratio}) shows that in the case of short
pump pulses all functional dependences of $R_{\rm short}$ and
$K_{\rm short}$ are identical and, numerically, the difference
between $R_{\rm short}$ and $K_{\rm short}$ is very small, $\sim
4.5\%$.

2) {\bf Long pulses}.

In the case of long pump pulses ($\eta\gg 1$) calculation of the
Schmidt number is similar to that described above for short
pulses, though the roles of the exponential (pump) and sinc-
functions are reversed: the pump spectral function is narrow and
the sinc-function is wide. Owing to this, the sinc-function in Eq.
(\ref{Freq-wave-function-2}) can be simplified in a way different
from that used in Eqs. (\ref{Delta-linear}), (\ref{K-linear}) and
(\ref{norm-linear}). In the case of long pulses the sum
$\nu_1+\nu_2$ is very small and the linear term in the argument of
the sinc-function can be dropped. As the result we can write down
Eqs. (\ref{K-general}) and (\ref{norm}) in the form
\begin{gather}
 \notag
 K^{-1}N^2=\int d\nu_1 d\nu_2 d\nu_1^\prime d\nu_2^\prime\\
 \notag
 \times\exp\left[-\frac{(\nu_1+\nu_2)^2\tau^2}{8\ln 2}\right]
 {\rm sinc}\left[\frac{LB}{2c}\frac{(\nu_1-\nu_2)^2}{\omega_0}\right]\\
 \notag
 \times\exp\left[-\frac{(\nu_1+\nu_2^\prime)^2\tau^2}{8\ln 2}\right]
 {\rm sinc}\left[\frac{LB}{2c}\frac{(\nu_1-\nu_2^\prime)^2}{\omega_0}\right]\\
 \notag
 \times\exp\left[-\frac{(\nu_1^\prime+\nu_2)^2\tau^2}{8\ln 2}\right]
 {\rm sinc}\left[\frac{LB}{2c}\frac{(\nu_1^\prime-\nu_2)^2}{\omega_0}\right]\\
  \label{K-long}
 \times\exp\left[-\frac{(\nu_1^\prime+\nu_2^\prime)^2\tau^2}{8\ln 2}\right]
 {\rm sinc}\left[\frac{LB}{2c}\frac{(\nu_1^\prime-\nu_2^\prime)^2}{\omega_0}\right]
\end{gather}
and
\begin{gather}
 \notag
 N=\int d\nu_1 d\nu_2\,\exp\left[-\frac{(\nu_1+\nu_2)^2\tau^2}{4\ln
 2}\right]\\
 \label{norm-long}
 \times\left\{{\rm  sinc} \left[\frac{LB}{2c}\frac{(\nu_1-\nu_2)^2}{\omega_0}\right]\right\}^2.
\end{gather}
The norm $N$ is calculated easily with $\nu_2=-\nu_1$ substituted
into the argument of the wide sinc$^2$-function, after which the
integral over $\nu_2$ in (\ref{norm-long}) takes the Gaussian form
to give
\begin{gather}
 \notag
 N=\frac{2\sqrt{\pi\ln 2}}{\tau}\int d\nu_1 \left[{\rm  sinc}
 \left(\frac{2LB\,\nu_1^2}{c\,\omega_0}\right)\right]^2\\
 =\frac{2\sqrt{\pi\ln 2}}{\tau}\sqrt{\frac{c\,\omega_0}{2LB}}\frac{4\sqrt{\pi}}{3},
 \label{norm-long-final}
\end{gather}
where the last factor on the right-hand side of Eq.
({\ref{norm-long-final}}) arises from integration over the
dimensionless variable $x=\nu_1\sqrt{2LB/c\,\omega_0}\,$:
$\int_{-\infty}^\infty\,dx\,{\rm sinc}^2(x^2)=4\sqrt{\pi}/3$.

The product of two exponential functions in the first and second
lines under the symbol of integrals in Eq. (\ref{K-long}) equals
to
\begin{gather}
 \notag
 \exp\left\{-\frac{\tau^2}{8\ln 2}\left[(\nu_1+\nu_2)^2+(\nu_1+\nu_2^\prime)^2\right]\right\}=\\
 \label{product}
 \exp\left\{-\frac{\tau^2}{4\ln 2} \left[\left(\nu_1+\frac{\nu_2+\nu_2^\prime}{2}\right)^2
 +\frac{(\nu_2-\nu_2^\prime)^2}{4}\right]\right\}.
\end{gather}
In dependence on $\nu_1$ this narrow function has a maximum at
$\nu_1=-(\nu_2+\nu_2^\prime)/2$, and this is the value that has to
be substituted instead of $\nu_1$ into the wide sinc-functions in
the first and second lines under the symbol of integrals in Eq.
(\ref{K-long}). This substitution makes the sinc-function
independent of $\nu_1$, and integration of the exponential
function (\ref{product}) over $\nu_1$ is easily performed.
Integration over $\nu_1^\prime$ in the last two lines of Eq.
(\ref{K-long}) is performed exactly in the same way. Altogether,
these two integrations (over $\nu_1$ and $\nu_1^\prime$) reduce
Eq. (\ref{K-long}) to a simpler form
\begin{gather}
 \notag
 K^{-1}N^2= \frac{4\pi\ln 2}{\tau^2}\int d\nu_2 d\nu_2^\prime
 \exp\left[-\frac{\tau^2}{8\ln 2} (\nu_2-\nu_2^\prime)^2\right]\times\\
 \label{K-long-semifinal}
 {\rm sinc}^2\left[\frac{LB}{8c\,\omega_0}(3\nu_2+\nu_2^\prime)^2\right]
 {\rm
 sinc}^2\left[\frac{LB}{8c\,\omega_0}(\nu_2+3\nu_2^\prime)^2\right],
\end{gather}
where the first and second sinc$^2$-functions arise,
correspondingly, from integration of the first and second pairs of
lines in Eq. (\ref{K-long}). In Eq. (\ref{K-long-semifinal}),
again, the exponential function is narrow and sinc-functions are
wide in their dependence on, e.g., $\nu_2^\prime$. Owing to this,
we can substitute $\nu_2^\prime$ by $\nu_2$ in the arguments of
the sinc-functions and take easily the arising Gaussian integral
over $\nu_2^\prime$ to get
\begin{gather}
 \notag
 K^{-1}N^2= \frac{8\sqrt{2}(\pi\ln 2)^{3/2}}{\tau^3}\,\int d\nu_2\,{\rm
 sinc}^4\left(\frac{2LB\nu_2^2}{c\,\omega_0}\right)\\
 \label{integration-fianal}
 =\frac{8\sqrt{2}(\pi\ln
 2)^{3/2}}{\tau^3}\sqrt{\frac{c\,\omega_0}{2LB}}
 \left[\frac{64}{105}(2^{3/2}-1)\sqrt{\pi}\right],
\end{gather}
where the last factor in square brackets is the exact expression
for the integral $\int_{-\infty}^\infty dx\, {\rm sinc}^4(x^2)$.

Eqs. (\ref{norm-long-final}) and (\ref{integration-fianal})
determine the Schmidt number in the long-pulse regime
\begin{gather}
 \notag
 K_{\rm
 long}=\frac{105\,\sqrt{\pi}}
 {72\,\sqrt{2\ln
 2}\,(2^{3/2}-1)}\frac{c\tau}{\sqrt{BL\lambda_0}}\\
 \label{K-long-final}
 =0.6 \,\frac{A\,\eta}{\sqrt{B}}\,\sqrt{\frac{L}{\lambda_0}}= 44\,\eta.
\end{gather}
Comparison of this result with that of Eq. (\ref{R-long}) shows
that in the case of long pulses, as well as in the case of short
pulses [Eqs. (\ref{ratio}) and (\ref{K-final})], the Schmidt
number and the parameter $R$ are rather close to each other.
Again, all functional dependences of Eq. (\ref{K-long-final})
and (\ref{R-long}) are identical. The parameters $K_{\rm long}$
and $R_{\rm long}$ differ only by numerical coefficients. In the
case of long pulses this difference is $\sim 20\%$, i.e., somewhat
larger than in the case of short pulses but, still, small enough
to consider $R$ as a good entanglement quantifier.

The Schmidt number for arbitrary pulse durations can be evaluated with the help of the
same square-root interpolation as in the case of $R$ [Eq.
(\ref{overall})]:
\begin{gather}
 \notag
 K(\eta)=\sqrt{K_{\rm short}^2+K_{\rm
 long}^2}\\
 \notag
 =\frac{A\,\eta}{\sqrt{B}}\,\sqrt{\frac{L}{\lambda_0}}\sqrt{\frac{(0.785)^2}{\eta}+(0.6\,\eta)^2}\\
 \label{K-overall}
 =\sqrt{\frac{(57.5)^2}{\eta}+(44\,\eta)^2}.
\end{gather}
The function $K(\eta)$ is plotted in Fig. \ref{Fig4} (the dashed
curve) together with $R(\eta)$. The difference between these two
curves is more pronounced in the case of long pump pulses and is
almost negligible in the case of short pulses. But even in the
case of long pulses the parameters $R$ and $K$ are close enough to
use the experimentally measurable parameter $R$ as the
entanglement quantifier.

The ratio of $R$ and $K$ parameters (\ref{overall}) and
(\ref{K-overall}) equals to
\begin{equation}
 \label{K-R-ratio}
 \frac{K(\eta)}{R(\eta)}\approx
 1.04\sqrt{\frac{1+0.586\,\eta^3}{1+\eta^3}}.
\end{equation}
Numerical coefficients on the right-hand side of this equation are
not related to any parameters of the medium or pump and arise form
$\ln 2$, $\sqrt{\pi}$ and similar factors. For this reason the
ratio $K/R$ can be considered as a universal function depending
only on the control parameter $\eta$ and valid for any crystals
(with the type-I phase matching) and any pump wave lengths.
Moreover, Eq. (\ref{K-R-ratio}) and all other results derived
above [except the very final numerical estimates in Eqs.
(\ref{ratio}), (\ref{R-long}), (\ref{R-short}), (\ref{K-final}),
(\ref{K-long-final}), and (\ref{K-overall})] can be applied to any
processes generating bipartite states of the form
(\ref{Freq-wave-function-2}) with arbitrary pairs of variables
substituting $\nu_1$ and $\nu_2$ and with any values and physical
origin of the constants $A$ and $B$. In particular, the results
described here can be applied directly to the earlier considered
angular entanglement of SPDC biphoton states in a continuous pump
laser \cite{we}. Specifically, for establishing one-to-one
correspondence between the spectral and angular entanglement, we
have to substitute frequencies $\nu_{1,\,2}$ by angles $\theta
_{1,\,2}$ determining directions of the photon wave vectors ${\bf
k}_{1,\,2}$. In accordance with the approach of Refs. \cite{we},
we can consider the angular divergence of the pump $\alpha_0$ as a
varying parameter substituting in the case of a monochromatic pump
the pump-pulse duration $\tau$. Then, to reduce the angular wave
function in the ``parallel geometry" of Refs. \cite{we} to the
form of Eq. (\ref{Freq-wave-function-2}), we have to substitute
the constants $A$ (\ref{A}) and $B$ (\ref{B}) and the control
parameter $\eta$ (\ref{eta}) by
\begin{equation}
 \label{angular}
 {\widetilde A}=\frac{\pi
 c}{\lambda_0}\frac{n_p^\prime}{n_p},\quad
 {\widetilde A}=\frac{\pi^2 c^2}{2n_p\lambda_0},\quad
 {\widetilde\eta}=\frac{4\ln 2\lambda_0}{\pi\alpha_0
 L}\frac{n_p}{n_p^\prime},
\end{equation}
where now $n_p^\prime$ is the angular derivative of the pump
refractive index $n_p$. With these substitutions done, we can use
all the results derived above for the Schmidt number $K$ and the
parameter $R$. In particular, with with the help of Eq.
(\ref{R-min}) we can find the minimal value of the parameter $R$
characterizing angular entanglement in the parallel geometry at
varying angular divergence of the pump $\alpha_0$:
\begin{equation}
 \label{r-min-angular}
 R_{\min\,\|}^{\rm angular}=
 n_p^\prime\,\sqrt{\frac{2L}{n_p\,\lambda_0}}.
\end{equation}
This result was not obtained earlier in refs \cite{we}, and it
shows that the degree of angular entanglement is high even in the
most unfavorable circumstances owing to the same large factor as
in the case of spectral entanglement, $\sqrt{L/\lambda_0}\gg 1$.

Finally, Eq. (\ref{K-R-ratio}) and the picture of Fig. \ref{Fig4} can
be used for direct experimental determining of the Schmidt number (if
the 20$\%$-accuracy evaluation of the degree of entanglement provided
by $R$ is insufficient). To find $K$ one has to measure $R$, to calculate
the value of the control parameter $\eta$ (\ref{eta}) corresponding to
experimental conditions, to find its location at the plot of Fig. \ref{Fig4}
and to find $K$. Alternatively, with known $R$ and $\eta$ the Schmidt number
is easily calculated with the help of Eq. (\ref{K-R-ratio}).

\section{Temporal structure of biphoton wave packets}
\subsection{General formulae}
Temporal and coordinate features of biphoton states are
characterized by the function ${\widetilde
\Psi}(t_1,\,t_2;\,z_1,\,z_2)$ depending on two times $t_1$ and
$t_2$ and two coordinates, $z_1$ and $z_2$,
\begin{gather}
 \notag
 {\widetilde \Psi}(t_1,\,t_2;\,z_1,\,z_2)=\int\,d\omega_1 d\omega_2\,
 \Psi(\omega_1,\,\omega_2)\\
 \label{Fourier}
 \times\exp\left\{-i\,\Big[\omega_1 t_1-k_1(\omega_1) z_1
 +\omega_2 t_2-k_2(\omega_2) z_2\Big]\right\}.
\end{gather}
The wave function ${\widetilde \Psi}$ determines the probability density
\begin{equation}
 \label{temp-prob-dens}
 \frac{dw}{dt_1dz_1dt_2dz_2}=|{\widetilde \Psi}(t_1,\,t_2;\,z_1,\,z_2)|^2
\end{equation}
of finding (registering) photons ``1" and ``2" in small vicinities
of arbitrary points (${z_1,\,t_1}$) and (${z_2,\,t_2}$) in
 the coordinate-time plane ($z,\,t$)
under the condition that $z_1$ and $z_2$ are within the crystal, i.e.,
$0\leq z_{1,\,2}\leq L$.

The exponential factors depending on $(z_1,\,t_1)$ and
$(z_2,\,t_2)$ on the right-hand side of Eq. (\ref{Fourier}) are
the time-coordinate wave functions of one-photon states with
momenta $k_1$ and $k_2$ (or frequencies $\omega_1$ and $\omega_2$)
\begin{gather}
 \notag
 e^{i(\omega_1t_1-k_1z_1)}=\langle
 z_1,\,t_1|k_1\rangle=
 \langle z_1,\,t_1|a^\dag_{k_2}|vac\rangle,\\
 \label{one-photon}
 e^{i(\omega_2t_2-k_2z_2)}=\langle z_2,\,t_2|k_1\rangle
 =\langle z_2,\,t_2|a^\dag_{k_2}|vac\rangle.
\end{gather}
The two-time and two-coordinate wave function of Eq.
(\ref{Fourier}) is the projection of the two-photon state vector
given by the second term on the right-hand side of Eq.
(\ref{qed-state}) upon the product of two coordinate states
\begin{gather}
 \notag
 {\widetilde \Psi}(t_1,\,t_2;\,z_1,\,z_2)=\\
 \langle z_1,\,t_1|\langle z_2,\,t_2|\sum_{\bf{k}_1,\,\bf{k}_2}\,\Psi(\bf{k}_1,\,\bf{k}_2)|\bf{k}_1,\,\bf{k}_2\rangle
 \label{qed-time-wf}
\end{gather}
for the case ${\bf k}_{1,2\;\perp}=0$.

Note that owing to the presence of coordinate parts in these
single-photon wave functions, transition from spectral to temporal
picture determined by Eq. (\ref{Fourier}) is not equivalent to
simple double Fourier transform of $\Psi(\omega_1,\,\omega_2)$
used in Refs. \cite{Keller-Rubin,LANL}. Though some results of
these two types of calculations can coincide (see Eq.
(\ref{Time-wf-at-exit-2}) and explanations below), in a general
case transition to the temporal picture is inseparable from
transition to the coordinate representation, and this is the
transformation determined by Eq. (\ref{Fourier}) that gives a
rigorously defined two-time and two-coordinate biphoton wave
function.

As it was done previously, the wave vectors $k_{1,\,2}(\omega_{1,\,2})$ in
Eq. (\ref{Fourier}) can be expanded in powers of small differences
$\nu_{1,\,2}=\omega_{1,\,2}-\frac{1}{2}\omega_0$, with only the first and
second orders to be retained
\begin{equation}
 \label{wave vectors}
 k_{1,\,2}(\omega_{1,\,2})\approx\frac{\omega_0}{2c}+\frac{\nu_{1,\,2}}{v_g^{(o)}}
 +\frac{2B}{c\,\omega_0}\,\nu_{1,\,2}^2 + ...
\end{equation}
By using this expansion, the definition of Eq. (\ref{Fourier}),
and the expression of Eq. (\ref{Freq-wf-int-dz}) for the
frequency-dependent wave function $\Psi(\omega_1,\,\omega_2)$, we
get the following integral representation for ${\widetilde \Psi}$
\begin{gather}
 \notag
 {\widetilde \Psi}(t_1,\,t_2;\,z_1,\,z_2)\propto\int_0^L dz\int\,d\nu_1
 d\nu_2\exp\left[-\frac{(\nu_1+\nu_2)^2\tau^2}{8\ln 2}\right]\\
 \notag
 \times\exp\Bigg[-i\nu_1\left(t_1-\frac{z_1}{v_g^{{(o)}}}\right)
 -i\nu_2\left(t_2-\frac{z_2}{v_g^{{(o)}}}\right)\\
 \notag
 +\,i\,\frac{A\,z}{c}\,(\nu_1+\nu_2)\Bigg]\\
 \label{Fourier-2}
 \times\exp\left[i\,\frac{B}{c\omega_0}\left(z-\frac{z_1+z_2}{2}\right)(\nu_1-\nu_2)^2
 \right],
\end{gather}
where, as previously, the squared frequencies
$\nu_{1,\,2}^2=\frac{1}{4}[\nu_1-\nu_2\pm(\nu_1+\nu_2)]^2$ in the
expansion (\ref{wave vectors}) are approximated by
$\frac{1}{4}(\nu_1-\nu_2)^2$ with the terms
$\pm\frac{1}{2}(\nu_1-\nu_2)(\nu_1+\nu_2)$ and $\frac{1}{4}(\nu_1+
\nu_2)^2$ dropped as giving only small corrections to other terms
$\propto\nu_1+\nu_2$ and $\propto(\nu_1+\nu_2)^2$ in the
exponential functions of Eq. (\ref{Fourier-2}).

In variables $u=\nu_1+\nu_2$ and $v=\nu_1-\nu_2$ the double
integral over $\nu_1$ and $\nu_2$ in Eq. (\ref{Fourier-2}) splits
for the product of two Gaussian integrals over $u$ and $v$ which
are easily taken to give
\begin{gather}
 \notag
 {\widetilde \Psi}(t_1,\,t_2;\,z_1,\,z_2)\propto\int_0^{\min(z_1,\,z_2)}
 \frac{dz}{\sqrt{\frac{z_1+z_2}{2}-z}}\\
 \notag
 \times\exp\left\{-\frac{2\ln
 2}{c^2\tau^2}\,\left[Az-\frac{c(t_1+t_2)}{2}+\frac{c(z_1+z_2)}{2v_g^{(o)}}\right]^2\right\} \times\\
 \label{Time-wave-function-2}
\exp\left\{\frac{i\pi
 c^2}{4B\lambda_0 (z_1+z_2-2z)}\left(t_1-t_2-\frac{z_1-z_2}{v_g^{(o)}}\right)^2\right\}
\end{gather}
Here we restricted the region of integration over $z$ by
$\frac{1}{2}(z_1+z_2)$ (instead of $L$) because we interpret $z$
as the coordinate of birth of a single-idler pair, and $z_1$ and
$z_2$ as possible locations of the emitted photons. As these
photons propagate to the direction of growing $z_{1,\,2}$, these
coordinates cannot be smaller than the coordinate of their birth,
i.e., $z$ must be smaller than $\min\{z_1,\,z_2\}$.

At the exit from the crystal, where $z_1=z_2=L$, Eq.
(\ref{Time-wave-function-2}) takes the form
\begin{gather}
 \notag
 {\widetilde \Psi}(t_1,\,t_2)\equiv{\widetilde \Psi}(t_1,\,L;\,t_2,\,L) \propto\int_0^L
 \frac{dz}{\sqrt{L-z}}\\
 \notag
 \times\exp\left\{-\frac{2\ln
 2}{\tau^2}\,\left[\frac{z}{v_g^{(e)}}+\frac{(L-z)}{v_g^{(o)}}-\frac{(t_1+t_2)}{2}\right]^2\right\} \\
 \label{Time-wf-at-exit}
 \times\exp\left\{i\frac{\pi
 c^2}{8B\lambda_0 (L-z)}(t_1-t_2)^2\right\}.
\end{gather}
Finally, by introducing new time variables
\begin{equation}
 \label{time-shift}
 {\widetilde t}_{1,\,2}=t_{1,\,2}-\frac{L}{v_g^{(o)}},\;
 t_+=\frac{1}{2}({\widetilde t}_1+{\widetilde t}_2),\;
 t_-={\widetilde t}_1-{\widetilde t}_2
\end{equation}
we can slightly simplify the first exponential factor on the
right-hand side of Eq. (\ref{Time-wf-at-exit}) and get
\begin{gather}
 \notag
 {\widetilde \Psi}({\widetilde t}_1,\,{\widetilde t}_2)\propto\int_0^L
 \frac{dz}{\sqrt{L-z}} \times\\
 \exp\left\{-\frac{2\ln
 2}{c^2\tau^2}\,(Az-c\,t_+)^2\right\}
 \label{Time-wf-at-exit-2}
 \exp\left\{i\frac{\pi
 c^2t_-^2}{8B\lambda_0 (L-z)}\right\}.
\end{gather}
The time shift $L/v_g^{(o)}$ in Eq. (\ref{time-shift}) is the time
during which the signal and idler photons moving exactly with
their group velocity cross all the crystal from 0 to $L$. Hence,
if $t_1=t_2=0$ is the time when the peak of the
temporal envelope of the pump enters the crystal, the zero point for
${\widetilde t}_{1,\,2}$ is the time when the signal and idler
photons, emitted at $t_1=t_2=0$ and moving with the speed
$v_g^{(o)}$, reach the end of the crystal.

Note that substitution of the integration variable $z=z^\prime+L$
and some additional shift of the time variables ${\widetilde t}_1$
and ${\widetilde t}_2$ reduce Eq. (\ref{Time-wf-at-exit-2}) to the
same form as that of Eq. (22) in the paper by Keller and Rubin
\cite{Keller-Rubin}. This equation of Ref. \cite{Keller-Rubin} was
obtained with the help of a simple double Fourier transformation
rather than via the transition from momentum to time-coordinate
representation determined by Eq. (\ref{Fourier}). Though seemingly
surprising, this coincidence is quite understandable. Indeed, in
terms of $z^\prime$, the crystal borders are $-L$ and $0$, and
at the exit from the crystal $z^\prime_1=z^\prime_2=0$. At this point,
the one-photon wave functions of Eq.
(\ref{one-photon}) (with $z$ substituted by $z^\prime$) turn into
$e^{-i\omega_{1,\,2} t_{1,\,2}}$, and the transformation of Eq.
(\ref{Fourier}) appears to be equivalent to the double Fourier
transformation. However, this is a very exceptional case related
to the assumed in Ref. \cite{Keller-Rubin} location of the crystal
at $(-L,0)$. Our attempt \cite{LANL} to apply a simple double
Fourier transformation to the case when the crystal location is
taken as $(0,L)$ gave the result qualitatively different from that
of Eq. (\ref{Time-wf-at-exit-2}) and, evidently, wrong. Here in
Eq. (\ref{Time-wf-at-exit-2}) we correct it with the help of
derivation based on the transformation from the frequency to
time-coordinate representation of Eq. (\ref{Fourier}).

Actually, the results of calculations have to be invariant with
respect to the choice of coordinates determining location of the
crystal. We can easily find that the transformation determined by
Eq. (\ref{Fourier}) provides such invariance whereas the simple
double Fourier transformation does not. Indeed, the difference
between Eq. (\ref{Time-wf-at-exit-2}) and Eq. (10) of Ref.
\cite{LANL} is in a different location of the singularity in the
integral over $z$: at $z=L$ (the exit edge of a crystal) in Eq.
(\ref{Time-wf-at-exit-2}) and $z=0$ (the entrance edge) in Eq.
(10) of Ref. \cite{LANL}. Location of the singularity is
determined by the factor in front of $(\nu_1-\nu_2)^2$ in the last
line of Eq. (\ref{Fourier-2}) and in similar equations of Refs.
\cite{Keller-Rubin} and \cite{LANL}. Let in a general case the
coordinates of the crystal edges be $z=L_1$ and $z=L_2=L_1+L$. We
are interested in the biphoton wave function at the exit edge,
i.e., $z_1=z_2=L_2$. In this case, the coefficient in front of
$(\nu_1-\nu_2)^2$ in the last line of Eq. (\ref{Fourier-2}) is
proportional to $z-L_2$, and the singularity in the final integral
over $z$ [similar to that of Eq. (\ref{Time-wf-at-exit-2})] occurs
at $z=L_2$, i.e., always at the exit edge of the crystal. In
contrast to this, in the scheme of calculation with the
double-Fourier transformation the coefficient in front of
$(\nu_1-\nu_2)^2$ in the last line of Eq. (\ref{Fourier-2})
becomes proportional to $z$, i.e., the singularity in the final
integral over $z$ occurs at $z=0$. This point can be arbitrarily
located with respect to the crystal and, evidently, the results
are not invariant with respect to the choice of the crystal
coordinates. The only case when the results of calculations in
these two approaches coincide is the case $L_1=-L$ and $L_2=0$,
and this is just the case of Ref. \cite{Keller-Rubin}.

\subsection{Short pump pulses}

\begin{figure}[h]
 \centering\includegraphics[width=8cm]{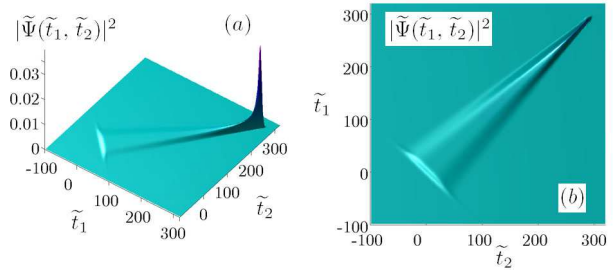}
\caption{{\protect\footnotesize {The function $|{\widetilde
\Psi}({\widetilde t}_1,\,{\widetilde t}_2)|^2$ for
$\tau=50\,{\rm fs}$, $(a)$ 3D plot and (b) the top view.}}} \label{Fig5}
\end{figure}

The probability density of registering photons at the exit from
the crystal at times ${\widetilde t}_1$ and ${\widetilde t}_2$ is
given by the squared absolute value of the wave function
${\widetilde \Psi}({\widetilde t}_1,\,{\widetilde t}_2)$
(\ref{Time-wf-at-exit-2}), determining a surface in the 3D space
$\Big\{{\widetilde t}_1,\,{\widetilde t}_2,\, |{\widetilde
\Psi}|^2\Big\}$. For short pump pulses $\tau=50{\rm fs}$ and for
the same values of all other parameters which were used above, the
surface ${\widetilde \Psi}({\widetilde t}_1,\,{\widetilde t}_2)$
is shown in Fig. \ref{Fig5}.

\begin{figure}[h]
 \centering\includegraphics[width=7cm]{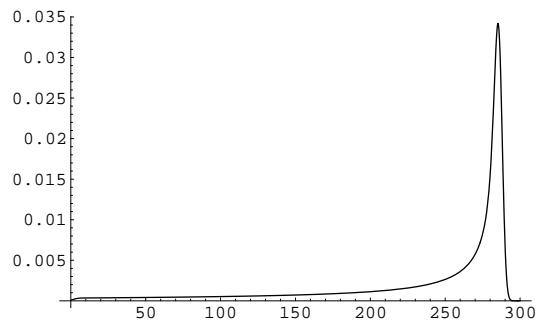}
\caption{{\protect\footnotesize {The diagonal squared wave
function $|{\widetilde \Psi}({\widetilde t}_1,\,{\widetilde
t}_1)|^2$.}}} \label{Fig6}
\end{figure}

The surface $|{\widetilde \Psi}({\widetilde t}_1,\,{\widetilde
t}_2)|^2$ is seen to have the form of a long plateau symmetric in
$t_1,\,t_2$. The plateau is rather wide in the beginning (at small
values ${\widetilde t}_{1,\,2}$) and is narrowing and heightening
with a growing value of ${\widetilde t}_1+{\widetilde t}_2$.
Around the point ${\widetilde t}_1={\widetilde t}_2=285.25$ the
plateau turns into a rather high and sharp peak. The diagonal
squared wave function $|{\widetilde \Psi}({\widetilde
t}_1,\,{\widetilde t}_1)|^2$ characterizes evolution of the
plateau-peak height in its dependence on ${\widetilde
t}_1+{\widetilde t}_2$, and this function is shown in Fig.
\ref{Fig6}. In accordance with the 3D picture
of Fig. \ref{Fig5}$\,(a)$, the curve of Fig. \ref{Fig6} switches
on rather quickly at $t=0$ (in a time on the order of the pulse
duration of the pump $\tau$), and then it has a long and slowly
growing plateau ending by a high and narrow peak at
$t=2.8525\,{\rm ps}$ followed by a very quick turn-off.

\begin{figure}[h]
 \centering\includegraphics[width=8.5cm]{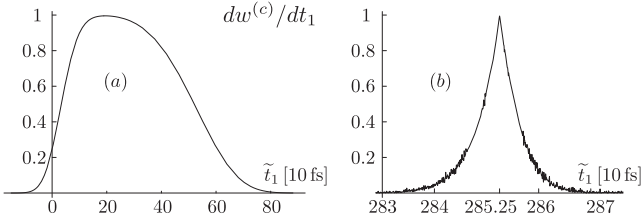}
\caption{{\protect\footnotesize {Coincidence signals at $(a)\,
{\widetilde t}_2=0$ and $(b)\,{\widetilde t}_2=2.8525,{\rm
ps}$.}}} \label{Fig7}
\end{figure}

The temporal shape and parameters of coincidence and
single-particle signals are defined in the same way as in the
cases of frequency or angular distributions. The coincidence
signal is determined by the probability density
$dw^{(c)}(t_1)/dt_1$ of registering at the exit surface of the
crystal a photon ``1" at a varying time ${\widetilde t}_1$ under
the condition that the photon ``2" of the same pair is registered
at the same place at some given time ${\widetilde t}_2$,
$dw^{(c)}(t_1)/dt_1\propto |{\widetilde \Psi}({\widetilde
t}_1,\,{\widetilde t}_2)|^2$ at ${\widetilde t}_2=const$. For
50-fs pulses of the pump, two examples of coincidence curves are
shown in Fig. \ref{Fig7}.
These curves correspond to the very beginning of the plateau and
to the peak of the surface $|{\widetilde \Psi}({\widetilde
t}_1,\,{\widetilde t}_2)|^2$ in Fig. \ref{Fig5}. The FWHM of the
curves $(a)$ and $(b)$ in Fig. \ref{Fig7} are equal to $\Delta
t_1^{(c)}({\widetilde t}_2=0)=486.3\,{\rm fs}$ and $\Delta
t_1^{(c)}({\widetilde t}_2=2.8525\,{\rm ps})=6\,{\rm fs}$. This
huge difference in FWHM shows clearly that in the temporal picture
the coincidence widths $\Delta t_1^{(c)}$ (or the duration of the
coincidence signal) depends very strongly on the registration time
$t_2$ of the second photon of a pair. This contrasts with the
spectral picture where $\Delta\nu_1^{(c)}$ was shown to be almost
independent of $\nu_2$. A strong dependence of $\Delta t_1^{(c)}$
on $t_2$ makes the temporal picture inappropriate for defining the
parameter $R$ and for determining in such a way the degree of
entanglement. Such a procedure can be used only in the spectral
picture. Some analogies can be found in the coordinate and
momentum pictures for massive particles, where spreading of wave
packets in the coordinate representation makes the latter
inconvenient for evaluation of the degree of entanglement via the
width-ratio parameter $R$. Whereas the momentum representation is
free from such a problem because momentum distributions do not
spread \cite{2004}. In the case of photons dispersion or
diffraction divergency plays the role analogous to spreading.

\begin{figure}[h]
 \centering\includegraphics[width=6.5cm]{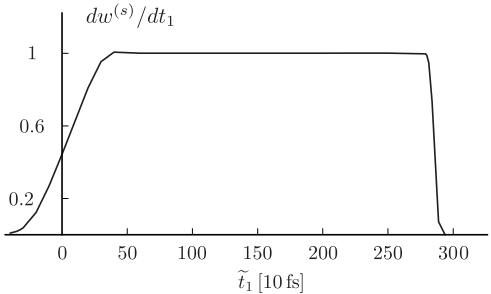}
\caption{{\protect\footnotesize {The normalized time-dependent
single-particle probability density.}}} \label{Fig8}
\end{figure}

The single-particle temporal density of probability to register a
photon by a single detector at a varying time $t_1$ is determined
by the integrated squared two-time wave function of Eq.
(\ref{Time-wf-at-exit-2}),
\begin{equation}
 \label{single-temporal}
 \frac{dw^{(s)}(t_1)}{dt_1}\propto\int dt_2 |{\widetilde\Psi}(t_1,\,t_2)|^2.
\end{equation}
For the pump-pulse duration as short as $\tau=50\,{\rm fs}$ the
result of numerical integration is plotted in Fig. \ref{Fig8}.
Duration of the single-particle signal, defined as FWHM of the
curve in Fig. \ref{Fig8}, equals to $2.837\,{\rm ps}$, which is
more than 50 times longer than the pump pulse.
Two other interesting features of the curve $dw^{(s)}(t_1)/dt_1$
are its constant value at a very long interval of time and the
anomalously long front wing (at small ${\widetilde t}_1)$.
Explanations are given below in the subsection {\bf D} on the
basis of simplified analytical formulas.

\subsection{Long pump pulses}
\begin{figure}[h]
 \centering\includegraphics[width=8.5cm]{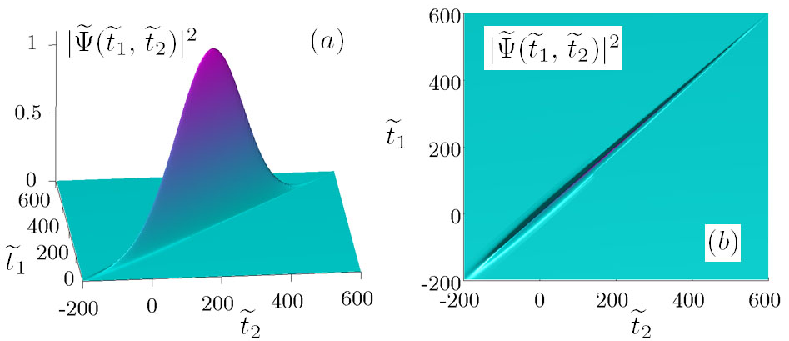}
\caption{{\protect\footnotesize {The function $|{\widetilde
\Psi}({\widetilde t}_1,\,{\widetilde t}_2)|^2$, 3D plot for
$\tau=2\,{\rm ps}$.}}} \label{Fig9}
\end{figure}

All the results shown in Figs. \ref{Fig5}-\ref{Fig8} correspond to
a short duration of the pump pulses ($\eta\ll 1$). In the case of
long pulses ($\eta\gg 1$) the picture changes rather
significantly. For $\tau=2\,{\rm ps}$ the 3D pictures of Fig
\ref{Fig5} are substituted by those of Fig. \ref{Fig9}.
The surface ${\widetilde\Psi}({\widetilde t}_1,\,{\widetilde
t}_2)$ is seen to have the form of a narrow long bump concentrated
along the diagonal ${\widetilde t}_1={\widetilde t}_2$  in the
plane (${\widetilde t}_1,\,{\widetilde t}_2$). As seen clearly in
Fig. \ref{Fig10}$\,(b)$, the width of the layer
${\widetilde\Psi}({\widetilde t}_1,\,{\widetilde t}_2)$ is almost
constant in the middle part of the bump, and the layer narrows
slightly at its ends. Quantitative characteristics of the temporal
photon distribution will be derived and presented in the last part
of the next Section.

\subsection{Analytical characterization}

{\bf a) Simplified formulae}. If the pump pulses are short enough,
$c\tau\ll AL$, the Eq. the first exponential factor in the integrand
of Eq. (\ref{Time-wf-at-exit-2})) is very narrow. Location of its
peak determines the coordinate $z_0=c\,t_+/A$,
a small vicinity of which gives the main contribution to the integral
over $z$. Under these conditions we can simplify the integrand by substituting
$z=z_0$ into the pre-exponential factor $1/\sqrt{L-z}$ and by
expanding $1/(L-z)$ in the second exponential factor in powers of
$|z-z_0|/z_0$: $1/(L-z)\approx 1/(L-z_0)+(z-z_0)/(L-z_0)^2$. As a
result of such simplifications, the integral in Eq.
(\ref{Time-wf-at-exit-2}) takes the Gaussian form with finite
limits of integration, which gives rise to the sum of two error
functions
\begin{gather}
 \notag
 {\widetilde\Psi}(t_+,\,t_-)\propto \frac{1}{\sqrt{LA-ct_+}}
 \exp\left[-a^2\left(\frac{ct_-}{LA-ct_+}\right)^4\right]\\
 \notag
 \times\Bigg\{ {\rm Erf}
 \left[\sqrt{2\ln 2}\frac{AL
 -ct_+}{c\tau}
 +i\,a\left(\frac{ct_-}{AL-ct_+}\right)^2\right]\\
 -{\rm Erf}\left[-\sqrt{2\ln 2}\,\frac{t_+}{\tau}
 +i\,a\left(\frac{ct_-}{LA-ct_+}\right)^2\right]\Bigg\},
 \label{erf}
\end{gather}
where
\begin{equation}
 \label{const-a}
 a=\frac{\pi c\tau A}{16\sqrt{2\ln 2}\,B\lambda_0}.
\end{equation}
The validity criterion of the approximation used for derivation of
Eq. (\ref{erf}) is $|z-z_0|\ll L-z_0$ with $z_0=ct_+/A$ and
$|z-z_0|\sim c\tau/A$, which gives
\begin{equation}
 \label{criterion}
 LA-ct_+\gg c\tau.
\end{equation}
As $t_+$ is associated with the location $z_0$ of the peak of the
pump envelope in the crystal, the condition (\ref{criterion})
means that the pump pulse is located far enough from the end of
the crystal, at a distance exceeding significantly $c\tau/A$.
Note, that the times $t_+=0$ and $t_+=LA/c$ are associated with
the pump pulse entry to and exit from the crystal.

If, in addition to (\ref{criterion}), the time $t_+$ or coordinate
$z_0$ obey the conditions $t_+\gg\tau$ or $z_0\gg c\tau/A$, the
pump envelope is located far from both edges of the crystal. In
this case the formula of Eq. (\ref{exp}) can be further
significantly simplified. Indeed, if a very short pump pulse is
located deep in the crystal, the limits of integration over $z$
can be extended to $-\infty$ and $+\infty$. This is equivalent to
approximation of the first and second error functions in Eq.
(\ref{erf}) by +1 and -1, and reduces finally Eq. (\ref{erf}) to
the simplest form:
\begin{gather}
 \label{exp}
 {\widetilde\Psi}(t_+,\,t_-)\propto \frac{1}{\sqrt{LA-ct_+}}
\exp\left[-a^2\left(\frac{ct_-}{LA-ct_+}\right)^4\right].
\end{gather}
Eqs. (\ref{erf}) and (\ref{exp}) are much more convenient than the
exact formula (\ref{Time-wf-at-exit-2}) for obtaining analytical
expressions for parameters characterizing coincidence and
single-particle signals.

{\bf b) Localization of the temporal biphoton wave
packet}.
\begin{figure}[h]
 \centering\includegraphics[width=8cm]{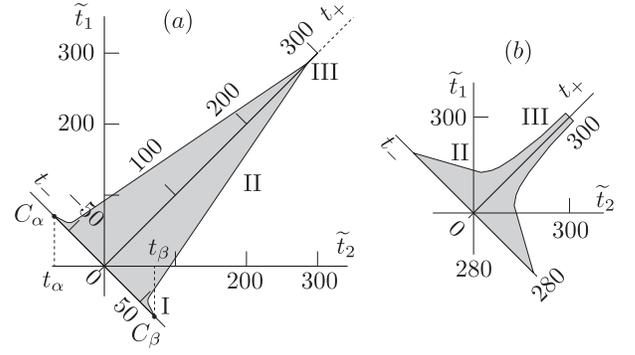}
\caption{{\protect\footnotesize {Localization region of the
function ${\widetilde\Psi}({\widetilde t}_1,\,{\widetilde t}_2)$;
${\widetilde t}_1,\,{\widetilde t}_2,\,t_+,\,{\rm and}\,t_-$ are
in units of $10^{-14}\,{\rm s}$; $(b)$ is the increased section
III of $(a)$.}}} \label{Fig10}
\end{figure}
The localization region of the wave function
${\widetilde\Psi}({\widetilde t}_1,\,{\widetilde t}_2)$ in the
(${\widetilde t}_1,\,{\widetilde t}_2$) plane can be defined in
the following way: we plot graphs of $|{\widetilde\Psi}|^2$ as a
function of $t_-$ at all given values of $t_+$ and find
coordinates $t_-^\pm(t_+)$ of these curves at half-maxima. Then
the functions $t_-^\pm(t_+)$ together with the lines $t_+=0$ and
and $t_+\approx 3\,{\rm ps}$ determine the borders of the
$|{\widetilde\Psi}|^2$ localization region. With the functions
$t_-^\pm(t_+)$ calculated numerically from the Eq.
(\ref{Time-wf-at-exit-2}) the result is given by the diagram of
Fig. \ref{Fig10}, in which $t_-^+(t_+)$ and $t_-^-(t_+)$ determine
the upper and lower borders of the shaded area. Defined in such a
way, the diagram of Fig. agrees perfectly with the
view-from-the-top of the surface ${\widetilde\Psi}({\widetilde
t}_1,\,{\widetilde t}_2)$ in Fig. \ref{Fig5}$\,(b)$. In Fig.
\ref{Fig10} one can distinguish clearly three regions of $t_+$, I,
II, and III, in which the dependence $t_-^\pm(t_+)$ is
significantly different. Below these regions are analyzed
separately.

{\bf c) Coincidence width in the region II}. The most extended of
three regions in Fig. \ref{Fig10} is the region of intermediate
values of $t_+$, II, where the wave function ${\widetilde
\Psi}(t_+,\,t_-)$ can be satisfactorily approximated by the
simplest exponential formula of Eq. (\ref{exp}). This equation
gives
\begin{equation}
 \label{localization}
 t_-^\pm(t_+)=\pm\frac{4}{c}\sqrt{\frac{\ln 2 AB\lambda_0}{\pi A
 c\tau}}\left(LA-ct_+\right).
\end{equation}
By definition, Eq. (\ref{localization}) determines the upper and
lower boundaries of the localization region in Fig. \ref{Fig10} at
a given value of $t_+$. However, the same equation can be used to
find the upper and lower boundaries of the localization region
${\widetilde t}_1^\pm$ at a given ${\widetilde t}_2$. To find
${\widetilde t}_1^\pm({\widetilde t}_2)$ one has to make a
substitution in Eq. (\ref{localization})
$t_+=\frac{1}{2}({\widetilde t}_1+{\widetilde t}_2)$, to replace
${\widetilde t}_-^\pm$ by ${\widetilde t}_1-{\widetilde t}_2$, and
to solve the arising equation with respect to ${\widetilde t}_1$.
In the first order in a small parameter $2\sqrt{\ln
2\,B\lambda_0/\pi c\tau}\approx 0.1\ll 1$ (at $\tau =50\,{\rm
fs}$) the solution is given by
\begin{equation}
 \label{localization-2}
 {\widetilde t}_1^\pm({\widetilde t}_2)={\widetilde
 t}_2\left(1\mp 4\sqrt{\frac{\ln 2 B\lambda_0}{\pi A
 c\tau}}\right)\pm\frac{4L}{c}\sqrt{\frac{\ln 2 AB\lambda_0}{\pi
 c\tau}}.
\end{equation}
From this equation we obtain the following analytical expression
for the width of the coincidence temporal distribution of photons
in the region II of Fig. \ref{Fig10}
\begin{equation}
 \label{coi-analyt}
 \Delta t_1^{(c)}({\widetilde t}_2)={\widetilde t}_1^+ -{\widetilde t}_1^-=
 \frac{8}{c}\sqrt{\frac{ B\lambda_0\ln 2}{\pi Ac\tau}}
 \left(LA-c{\widetilde t}_2\right).
\end{equation}
This expression shows that spread of photons coming in pairs to
the exit from crystal over time of their arrival is determined by
a joint influence of the temporal walk-off and dispersion. The
spread is large at small values of the observation time
characterized by ${\widetilde t}_2$ and falls with a growing
${\widetilde t}_2$.

{\bf d) The region I}. In the region I the simplest approximate
formula of Eq. (\ref{exp}) is invalid. But the formula of Eq.
(\ref{erf}) works perfectly and can be used to find analytical
expressions determining both the shape of boundaries of
localization region in Fig. \ref{Fig10} and the coincidence width
of the temporal distribution of photons. As in the region I the
time $t_+$ is very small, the first error function in Eq.
(\ref{erf}) equals unity, and the term $t_+$  can be dropped
everywhere except the first term in the argument of the second
error function to give
\begin{equation}
 \label{small t+}
 {\widetilde\Psi}\propto e^{-y^4}[1+{\rm Erf}(b+i\,y^2)],
\end{equation}
where $b=\sqrt{2\ln 2}\,t_+\tau$, $y=ct_-\sqrt{a}/LA$, and $a$ is
defined in Eq. (\ref{const-a}). From Eq. (\ref{small t+}) we find
easily (e.g., with the help of simple calculation in
``Mathematica") that at $b=0$ the HWHN (half-width at the
half-maximum) of $|{\widetilde\Psi}|^2$ equals one, and at growing
but small values of $b$ it falls as $1-0.24\, b$. In terms of
$t_+$ and $t_-$ this yields
\begin{gather}
 \notag
 t_-^\pm(t_+)=\pm\frac{LA}{c\sqrt{a}}\left(1-0.24\frac{\sqrt{2\ln 2}\,t_+}{\tau}\right)\\
 \label{t+&-inI}
 =\frac{4L(2\ln 2)^{1/4}\sqrt{AB\lambda_0}}{c\sqrt{\pi c\tau}}
 \left(1-0.24\frac{\sqrt{2\ln 2}\,t_+}{\tau}\right).
\end{gather}
The lines $t_-^\pm(t_+)$ determined by Eq. (\ref{t+&-inI}) are
sharper than the lines, determined by Eq. \ref{localization}). The
crossing points of these  lines occur at $t_+\approx 43\,{\rm fs}$
and they determine the boundary between the zones I and II in Fig.
\ref{Fig10}. As seen well in Fig. \ref{Fig10}, the width of the
localization region in the $t_-$-direction is maximal at $t_+=0$
where, in accordance with Eq. (\ref{t+&-inI}), it equals to
\begin{gather}
 \notag
 \Delta t_{-\, \max}=t_-^+(t_+=0)-t_-^-(t_+=0)\\
 \label{zero t+ width in t-}
 =\frac{8L(2\ln 2)^{1/4}\sqrt{AB\lambda_0}}{c\sqrt{\pi c\tau}}\approx 1.46\,{\rm ps}.
\end{gather}

{\bf e) Pulse shape of a single-particle signal}. A large width of
the wave-packet localization region in the $t_-$-direction at
$t_+=0$ explains the origin and parameters of the anomalously long
front wing of the single-particle signal shown in Fig. \ref{Fig8}.
Indeed, owing to the  condition $t_+=0$, in the (${\widetilde
t}_1,\,{\widetilde t}_2$) frame the points $C_\alpha$ and
$C_\beta$ in Fig. \ref{Fig10} have coordinates ${\widetilde
t}_{1\,\alpha}=-\,{\widetilde t}_{2,\,\alpha}=\frac{1}{2}\Delta
t_{-\, \max}$ and ${\widetilde t}_{1\,\beta}=-\,{\widetilde
t}_{2,\,\beta}=-\frac{1}{2}\Delta t_{-\, \max}$, where $\Delta
t_{-\, \max}$ is given by Eq. (\ref{zero t+ width in t-}). As the
$t_1$-dependent single-particle signal can be considered as the
sum of all coincidence signals at all values of $t_2$, it's clear
that at ${\widetilde t}_2<{\widetilde t}_{2\,\alpha}$ there are no
signals at all, either in coincidence or single-particle
measurements. Hence, ${\widetilde t}_{2\,\alpha}$ is that tim
since which the single-particle signal become detectable.
Actually, ${\widetilde t}_{2\,\alpha}$ is the beginning of the
transient period during which the amplitude of the single-particle
grows. This transient period ends at ${\widetilde t}_2={\widetilde
t}_{2\,\beta}$, and its duration equals to $\Delta t_{-\, \max}$
(\ref{zero t+ width in t-}). The same value characterizes the
duration $\tau_{\rm fw}$ of the front wing of the single particle
signal considered as a function of $t_1$:
\begin{equation}
 \label{front-wing}
 \tau_{\rm fw}={\widetilde t}_{1\,\alpha}-{\widetilde
 t}_{1\,\beta}=
 \frac{8L(2\ln 2)^{1/4}\sqrt{AB\lambda_0}}{c\sqrt{\pi c\tau}}.
\end{equation}

Another interesting feature of the single-particle signal is its
constant value during all the its duration after ending of the
initial transient period (Fig. \ref{Fig8}). This feature is easily
explained by Eq. (\ref{exp}) and (\ref{coi-analyt}). Indeed, at a
given ${\widetilde t}_1$ the single-particle yield
(\ref{single-temporal}) can be estimated approximately as the
product of the hight $|{\widetilde\Psi}({\widetilde
t}_1,\,{\widetilde t}_1)|^2$ and width $\Delta
t_2^{(c)}({\widetilde t}_1)$ of $|{\widetilde\Psi}{\widetilde
t}_1,\,{\widetilde t}_2)|^2$ considered as a function ${\widetilde
t}_2$. The first of these two factors is given by Eq. (\ref{exp})
with $t_-=0$ and, evidently, $|{\widetilde\Psi}({\widetilde
t}_1,\,{\widetilde t}_1)|^2\propto 1/(LA-c{\widetilde t}_1)$. As
for the width $\Delta t_2^{(c)}({\widetilde t}_1)$, it is given by
Eq. (\ref{coi-analyt}) with ${\widetilde t}_2$ on its right-hand
side substituted by ${\widetilde t}_1$, and this gives $\Delta
t_2^{(c)}({\widetilde t}_1)\propto LA-c{\widetilde t}_1$. Hence,
with a growing ${\widetilde t}_1$ the peak hight of the function
$|{\widetilde\Psi}({\widetilde t}_1,\,{\widetilde t}_2)|^2$ (in
its dependence on ${\widetilde t}_2$) grows, whereas its width
falls, but their product remains constant (independent of
${\widetilde t}_1$), i.e., the dependence $dw^{(s)}({\widetilde
t}_1)/d{\widetilde t}_1$ has the form of a plateau.

Finally, the total duration of the single-particle signal is
determined by the condition that the real part of the argument of
the second error-function in Eq. (\ref{erf}) changes its sign,
i.e.
\begin{equation}
 \label{single-duration}
 \tau_{\rm single}\approx \frac{LA}{c}
 =\frac{L}{{\rm v}_g^{(e)}}-\frac{L}{{\rm v}_g^{(o)}}.
\end{equation}
This result is absolutely understandable qualitatively: $\tau_{\rm
single}$ is the period of time between the arrival to exit from
the crystal of almost first and almost last observable SPDC
photons or the delay time between the arrival of the first
ordinary-wave photons and of the pump (extraordinary wave).

{\bf f) The region III in the diagram of Fig. \ref{Fig10}}. This
is the region behind the peak of the the photon distribution in
Fig. \ref{Fig5}$\,(a)$, where the amount of photons is vary small.
Nevertheless, this region is interesting, because here formation
of the localization region of the temporal wave packet is related
to a mechanism qualitatively different from those described above
for the regions I and II. In the region III the condition of Eq.
(\ref{criterion}) is not fulfilled, the fraction $1/(L-z)$ in the
integrand of Eq. (\ref{Time-wf-at-exit-2}) cannot be expanded in
powers of $(z-z_0)/(L-z_0)$, and the simple formulas of Eqs.
(\ref{erf}) and (\ref{exp}) are invalid. But some estimates can be
done directly on the ground of Eq. (\ref{Time-wf-at-exit-2}).
Indeed, from the second exponential factor in the integrand on the
right-hand side of Eq. (\ref{Time-wf-at-exit-2}) we can deduce the
following estimate of the characteristic half-width of the
wave-packet localization region
\begin{equation}
 \label{Dt-III}
 \Delta t_-\pm\sim \pm\frac{\sqrt{B\lambda_0(L-z)}}{c}.
\end{equation}
If the peak of the laser pulse is located exactly at the end of
the crystal, $ct_+=LA$, then the maximal deviation of $z$ from
$L$, where the integrand of Eq. (\ref{Time-wf-at-exit-2}) is not
small, is $L-z\sim c\tau/A$. Substitution of this expression into
Eq. (\ref{Dt-III}) gives
\begin{equation}
 \label{Dt-III-2}
 \Delta t_-^\pm\sim \pm\frac{\sqrt{B\lambda_0\tau}}{\sqrt{cA}}.
\end{equation}
This expression corresponds to the very beginning of the zone III
in Fig. \ref{Fig10}.

If $ct_+>L$, the peak of the pump pulse is located behind the
crystal and on a small part if its rear wing remains in the
crystal. By expanding the Gaussian exponent in powers of $L-z$, we
find that now $L-z\sim c^2\tau^2/{[A(ct_+-LA)]}$. By substituting
this expression into Eq. (\ref{Dt-III}), we find the law by which
the zone III in Fig. \ref{Fig10} slowly narrows
\begin{equation}
 \label{Dt-III-3}
 \Delta t_-^\pm\sim \pm\tau\sqrt{\frac{B\lambda_0}{A(ct_+-L)}}.
\end{equation}

\begin{figure}[h]
 \centering\includegraphics[width=6cm]{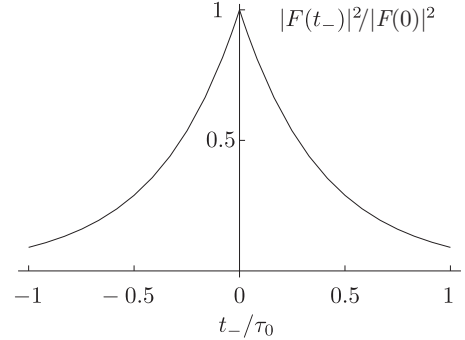}
\caption{{\protect\footnotesize {Squared absolute value of the expression on
 the right-hand side of Eq. (\ref{F(t-)-final}).}}} \label{Fig11}
\end{figure}

{\bf g) Long pulses}. The regime of long pulses of the pump arises
if the pulse duration is so long that $\tau\gg LA/c$ or $\eta\gg
1$ where $\eta$ is given by Eq. (\ref{eta}). In this case the term
proportional to $z$ can be dropped in the argument of the first
exponential function in Eq. (\ref{Time-wf-at-exit-2}) to give
\begin{equation}
 \label{Time-wf-long}
 {\widetilde \Psi}({\widetilde t}_1,\,{\widetilde t}_2)\propto
 \exp\left[-2\ln 2\left(\frac{t_+}{\tau}\right)^2\right]F(t_-),
\end{equation}
where
\begin{equation}
 \label{F(t-)-long}
 F(t_-)=\int_0^L
 \frac{dz}{\sqrt{L-z}}\exp\left[i\frac{\pi
 c^2t_-^2}{8B\lambda_0 (L-z)}\right].
\end{equation}
Eq. (\ref{Time-wf-long}) shows that for long pulses the
dependences of the wave function ${\widetilde\Psi}$ on $t_+$ and
$t_-$ are factorized, and the dependence of ${\widetilde\Psi}$ on
$t_+$ repeats the time-dependence of the pump envelope. As for the
integral over $z$ in Eq. (\ref{F(t-)-long}) determining the
dependence of ${\widetilde\Psi}$ on $t_-$, it can be calculated to
expressed $F(t_-)$ in terms of the error function. To show this,
we have to substitute the variable of integration
$z=L\left[1-\frac{i}{s}\left(\frac{t_-}{\tau_0}\right)^2\right]$,
where
\begin{equation}
 \label{tau-0}
 \tau_0=\frac{1}{c}\sqrt{\frac{8B\lambda_0L}{\pi}}
\end{equation}
and $s$ is a new purely imaginary variable. Then the integral
(\ref{F(t-)-long}) takes the form
\begin{equation}
 \label{F(t-)-s}
 F(t_-)\propto\frac{|t_-|}{\tau_0}\int_{i(t_-/\tau_0)^2}^{i\infty}\frac{ds\,e^{-s}}{s^{3/2}}.
\end{equation}
In the complex plane $s$ the integration contour can be turned
clockwise for $90^\circ$ to  be made parallel to the real axis. At
this new integration contour
$s=i\left(\frac{t_-}{\tau_0}\right)^2+s^\prime$, where
$s^\prime={\rm Re}(s)$ is a new real integration variable changing
from 0 to $\infty$:
\begin{gather}
 \notag
 F(t_-)\propto\frac{|t_-|}{\tau_0}\int_0^{\infty}
 \frac{ds^\prime\,e^{-s^\prime}}
 {\left[i\left(\frac{t_-}{\tau}\right)^2+s^\prime\right]^{3/2}}\\
 \label{F(t-)-final}
 = 1+\sqrt{i\pi}\exp\left[i\left(\frac{t_-}{\tau_0}\right)^2\right]\frac{t_-}{\tau_0}
 \left[-1+{\rm Erf}\left(\sqrt{i}\,\frac{|t_-|}{\tau_0}\right)\right],
\end{gather}
where, as usual, all constant and phase factors are dropped.
Correctness of the last transformation in Eq. (\ref{F(t-)-final})
(transition to the error function) can be checked, e.g., by direct
calculation of the integral in ``Mathematica". The normalized
squared absolute value of $F(t_-)$ (\ref{F(t-)-final}) is plotted
in Fig. \ref{Fig11} as a function of the dimensionless variable
$t_-/\tau_0$.
As the dependence of the function $F(t_-)$ on all parameters of
the crystal and pump is concentrated only in the scaling factor
$\tau_0$ (\ref{tau-0}), the curve of Fig. \ref{Fig11} is universal
and valid for all values of these parameters under the only
condition that pump pulses are long, $\tau\gg LA/c$. In
particular, this curve and the analytical results of Eqs.
(\ref{Time-wf-long}) and (\ref{F(t-)-final}) are perfectly valid
for the above considered case of 2-ps long pulses, for which the
temporal wave packets are described in Fig. \ref{Fig9}.

Found from Eqs. (\ref{Time-wf-long}) and (\ref{F(t-)-final})
sigle-particle and coincidence widths of the temporal distribution
of photons are given by
\begin{equation}
 \label{width-time-long}
 \Delta t_1^{(s)}=\tau\; {\rm and}\; \Delta
 t_1^{(c)}=0.555\tau_0=\frac{0.555}{c}\sqrt{\frac{8BL\lambda_0}{\pi}},
\end{equation}
where 0.555 is the FWHM of the curve in Fig. \ref{Fig11}. In
contrast to the case of short pulses, in the regime of long pulses
of the pump wave the coincidence width $\Delta
 t_1^{(c)}$ does not depend on $t_2$. For these reason, Eqs.
 (\ref{width-time-long}) can be used for determining the entanglement
parameter $R_t$ in the temporal representation:
\begin{equation}
 \label{R-t}
 R_t=\frac{\Delta t_1^{(s)}}{\Delta
 t_1^{(c)}}=\frac{c\tau}{0.555}\sqrt{\frac{\pi}{8BL\lambda_0}}\approx
 0.75 R_{\rm long}\approx 0.94 K_{\rm long},
\end{equation}
where $R_{\rm long}$ is the $R$-parameter (\ref{R-long})
calculated for long pulses in the frequency representation and
$K_{\rm long}$ (\ref{K-long}) is the above-calculated Schmidt
number in the long-pulse asymptotic regime. As seen from Eq.
(\ref{R-t}), the temporal $R$-parameter, $R_t$, has all the same
functional dependences as $R_{\rm long}$ and $K_{\rm long}$ and
slightly differs from them in numerical coefficients. Moreover,
$R_t$ appears to be strikingly close to the Schmidt number $K_{\rm
long}$, and this confirms once again a good quality of the
$R$-parameter as the entanglement quantifier.

The coincidence width $\Delta t_1^{(c)}$  (\ref{width-time-long}),
or equal to it ``scaling factor"
$\tau_0$ (\ref{tau-0}), can be interpreted as the correlation time of
photons in the biphoton pair. It's important to emphasize that this time
does not depend on the pump-pulse duration $\tau$. Hence, the
correlation time remains finite even in the case of an infinitely long
pulses or a purely monochromatic pump. The only reason for a finite time
of correlation in the case of long pump pulses is the dispersion. If
the dispersion constant $B$ formally tends to zero, the correlation time
$\tau_0$ tends to zero too.

Note also that, owing to the assumption that pump pulses are long,
in this limit the temporal walk-off appears to be completely
switched off. The scaling duration $\tau_0$ (\ref{tau-0}), as well
as the width and all structure of the curve in Fig. \ref{Fig11}
are determined only by the dispersion constant $B$, rather than by
any combination of $B$ and $A$ as it was shown to occur in the
case of short pulses. Finally, the structure of the curve in Fig.
\ref{Fig9} reminds quite strongly the structure of the coincidence
curves for short pulses in the region of the peak [Fig.
\ref{Fig7}$\,(b)$] and in the zone III of Fig. \ref{Fig10}. This
resemblance is related to the increasing role of dispersion and
decreasing role of the temporal walk-off in the cases when a
short-pulse pump exits the crystal.

\section{Conclusion}

Let us summarize here our main results.

1. For the degenerate collinear type-I
SPDC process the parameter $R$, characterizing the degree of spectral
entanglement,  is found and shown to be large at any values of
pulse duration $\tau$ of the pump. Even in the most unfavorable conditions
$\tau\sim 1\,{\rm ps}$, where $R(\tau)$ has a minimum, its value does
not fall below 70, i.e., $R(\tau)\geq R_{\min}\gg 1$.

2. The control parameter $\eta$ (\ref{eta}) separating the regions
of short and long pulses is found to be determined  by the ratio
of the double pump-pulse duration to the difference of times
during which a crystal is crossed by the pump and idler/signal
photons.

3. By using fundamental features of the biphoton spectral wave
function we managed to find the Schmidt number $K$ in the cases of
short and long pulses (compared to 1 ps), though the wave function
was taken in the form significantly different from the
double-Gaussian one. By comparing $R$ and $K$, we found that they
are very close to each other. All functional dependences of $R$
and $K$ are identical, and the difference in numerical
coefficients does not exceed $20\,\%$. For the class of bipartite
wave functions we consider, we found a simple universal formula
[Eq. (\ref{K-R-ratio})] establishing relation between $R$ and $K$
for any values of the control parameter $\eta$. This means that as
soon as the parameter $R$ is measured and $\eta$ is evaluated, the
Schmidt number can be easily found too.

4. In the temporal picture, a structure of the two-time biphoton
wave packet $|\Psi({\widetilde t}_1,\,{\widetilde t}_2)|^2$ is
investigated both numerically and analytically (the latter - with
the help of a series of very simple formulas we have derived), in
both cases of short and long pump pulses. It's found that in these
two cases (short and long pump pulses) the shapes and localization
regions of the temporal biphoton wave packet are significantly
different. In the case of long pump pulses the localization region
of $|\Psi({\widetilde t}_1,\,{\widetilde t}_2)|^2$ is more or less
usual: it looks like a long and thin cigar elongated along the
diagonal ${\widetilde t}_1={\widetilde t}_2$ in the plane
$({\widetilde t}_1,\,{\widetilde t}_2)$ (Fig. \ref{Fig9}$(b)$). In
contrast to this, in the case of short pump pulses, the
localization region of $|\Psi({\widetilde t}_1,\,{\widetilde
t}_2)|^2$ is found to be wedge-shaped (Figs. \ref{Fig5} and
\ref{Fig10}). Again, this ``wedge" is elongated along the diagonal
${\widetilde t}_1={\widetilde t}_2$. But, very unusual, its width
in the perpendicular direction (${\widetilde t}_1=-{\widetilde
t}_2$) changes in very wide limits. In our example used above for
estimates, from about 1.5 ps in a wide part of the ``wedge" down
to a few femtoseconds in the narrow part.

5. The above-described structure of the temporal wave packet in
the case of short pump pulses gives rise to the following specific
features of the single-particle signals. $(a)$ Duration of the
single-particle signal is much longer than that of the pump
($\sim$3 ps in our example compared to $\tau =50$ fs). $(b)$ the
front wing of the single-particle signal is anomalously long,
$\sim$ 1 ps, whereas the rear wing is as short as the pump pulse
itself. $(c)$ In its middle part, the envelope of the
single-particle signal has a form of a plateau, i.e. is constant.
The long duration of the single-particle signal is related to the
temporal walk-off effect only, whereas its plateau shape and a
long front wing are related to a combined action of the temporal
walk-off and dispersion.

6. In the same case of short pump pulses, coincidence signals have
duration which depends strongly on the observation time. Soon
enough after the first signal/idler photons arrival to the end of
the crystal, duration of coincidence signals (equal to the
characteristic time of photon correlation in a pair) is very long,
($\sim$ 1 ps), whereas close to the end of observability the
duration of coincidence signals correlation time become very short
(a few fs). This uncertainty in the definition of the coincidence
duration makes the temporal picture in the case of short pump
pulses very inconvenient and inappropriate for defining the
$R$-parameter and evaluating in such a way the degree of
entanglement of a biphoton state.

7. In the case of long pulses the wave packet $|{\widetilde\Psi}({\widetilde t}_1,\,{\widetilde
t}_2)|^2$ is concentrated along the diagonal ${\widetilde t}_1={\widetilde t}_2$ and is very
narrow and long. Duration of coincidence signals is small,
determined purely by dispersion, and independent of the observation time,
i.e. constant. Owing to the last reason, the parameter $R$ can be
clearly and unambiguously defined, and it appears to be very close
to the Schmidt number. Actually, a high degree of closeness
between the parameters $R$ and $K$ repeats itself symmetrically in
the cases of $(a)$ short pulses and frequency representation and
$(b)$ long pulses and temporal representation. In both cases, with
a good accuracy, $R=K\gg 1$.

\section*{Acknowledgement}
We are grateful to  S.P. Kulik and J.H. Eberly for stimulating
discussions. The work was supported partially by the RFBR grants
05-02-16469 and 08-02-01404, the Dynasty Foundation and the
Russian Federation President's Grant MK1283.2005.2.

\end{document}